\pdfoutput=1
\documentclass{JINST}

\newcommand{\subsubsubsection}[1]{\paragraph{#1}\mbox{}\\}
\usepackage{textcomp}
\usepackage{float}
\usepackage{ifpdf}

\title{Prototype electronics for the silicon pad layers of the future Forward Calorimeter (FoCal) of the ALICE experiment at the LHC}

\author{O.~Bourrion\thanks{Corresponding author.}, D.~Tourres, R.~Guernane, C.~Arata, J.-L.~Bouly, N.~Ponchant.\\
Univ. Grenoble Alpes, CNRS, Grenoble INP\textsuperscript{$\dagger$}, LPSC-IN2P3, 38000 Grenoble, France \\
$\dagger$ Institute of Engineering Univ. Grenoble Alpes
}


\abstract{
A Forward Calorimeter (FoCal) has been proposed as part of the ALICE upgrades for data taking from 2029 onwards.
The FoCal will feature a sampling electromagnetic calorimeter segmented into 110 towers supplemented by a hadron calorimeter.
The electromagnetic calorimeter will be composed of 20 passive layers of tungsten absorber interleaved with 18 active layers of low-granularity silicon pad sensors and two layers of high-granularity pixel detectors.
Each pad layer will be read out by 110 silicon pad sensors of 72 channels, amounting to a total of 1980 sensors.

This paper describes, from front-end to back-end, the electronics developed to instrument a tower prototype composed of 18 silicon pad sensors as well as a design proposal for the full-detector readout system.
}

\begin{document}

\section{Introduction}

A Forward Calorimeter (FoCal) has been proposed as part of the ALICE upgrades in the Large Hadron Collider (LHC) Long Shutdown 3 (LS3) for data taking from 2029 onwards \cite{focal_loi}.
FoCal extends the scope of the ALICE experiment, originally designed for a comprehensive study of hot and dense partonic matter, to include new capabilities for the study of the small-$x$ partonic structure of nucleons and nuclei.

FoCal provides unique opportunities to study parton distribution functions (PDFs) in the hitherto unexplored Bjorken-$x$ region down to $x\sim 10^{-6}$ and low momentum transfer $Q^2\approx 4\,\mathrm{GeV}/c$, where PDFs are expected to evolve nonlinearly due to high gluon densities, potentially leading to saturation.
The main goal of FoCal is to provide high-precision inclusive measurements of direct photons and jets, as well as measurements of gamma–jet and jet–jet coincidences, in pp and p–Pb collisions, and $\mathrm{J}/\psi$ photoproduction in ultraperipheral Pb–Pb collisions.

This article reports on the electronics development carried out for the readout system of the FoCal electromagnetic calorimeter.
The proposed solution was driven by the requirement to sustain the running scenario expected at ALICE for the high luminosity LHC (HL-LHC) phase \footnote{Radiation doses corresponding to integrated luminosities of $50\ \mathrm{nb}^{-1}$ of p–Pb, and $6\ \mathrm{pb}^{-1}$ of pp collisions and $1\ \mathrm{MHz}$ interaction rate.}.   

\section{The FoCal detector design}

\begin{figure}[hbtp]
\centering
\includegraphics[angle=0,width=0.7\textwidth]{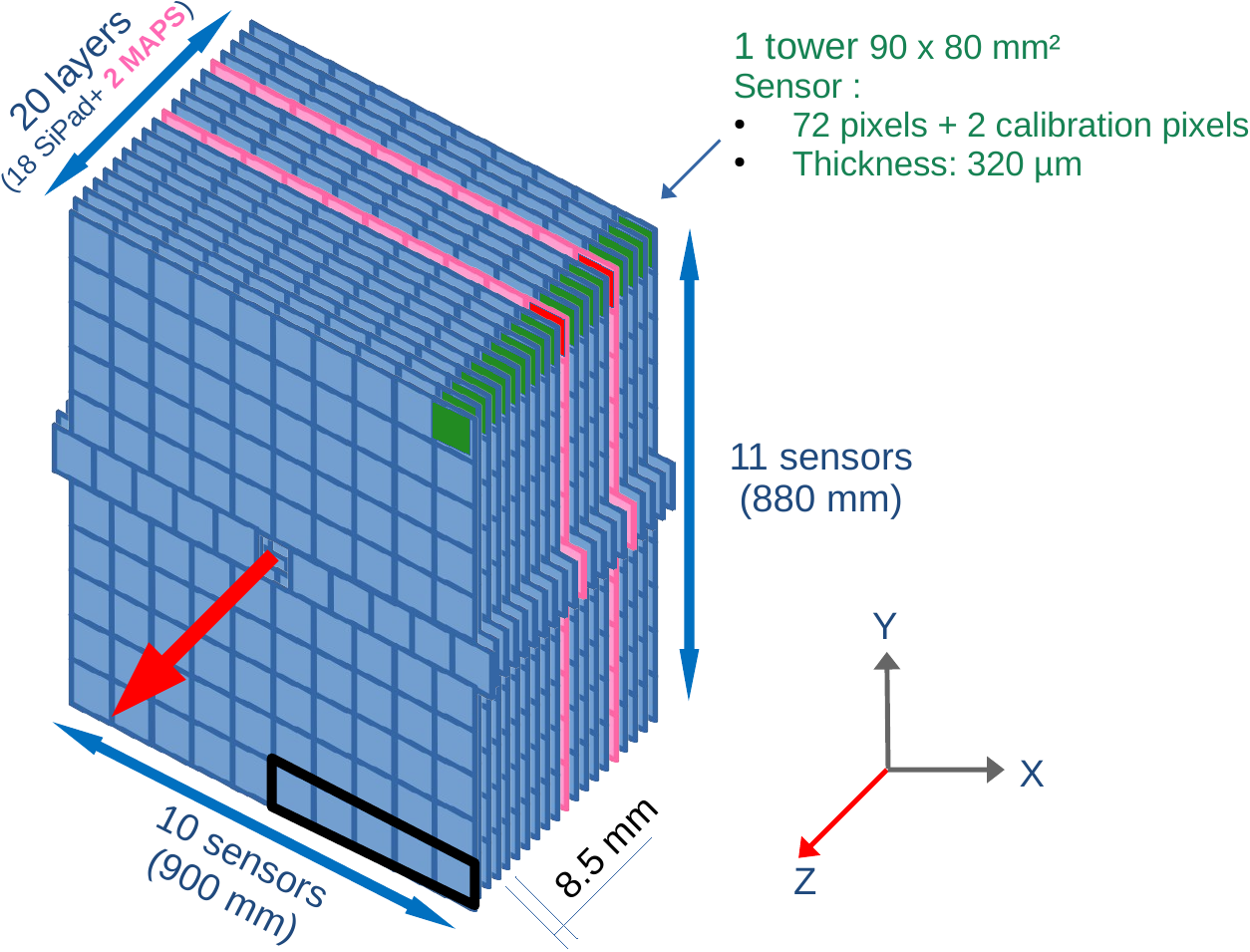}
\caption{
\label{detOverview} 
Longitudinal structure of the FoCal electromagnetic calorimeter. The ALICE coordinate system is shown as defined in \cite{ALICE_numbering} with the beam direction along the $z$ axis depicted by the red arrow. The FoCal electromagnetic calorimeter has a total of 20 layers with an $8.5\,\mathrm{mm}$ pitch along the $z$-axis: 18 layers of tungsten and silicon pads (blue) with low granularity ($\sim 1\,\mathrm{cm^2}$) interleaved with two layers of tungsten and silicon pixels MAPS (pink) with high granularity \cite{Alpide_paper}.
The pad layers are read out by 110 silicon pad sensors organized in an array of $10 \times 11$ sensors.
Each sensor has a thickness of 320\,\textmu m and features 72 channels and 2 calibration cells.
}
\end{figure}

The FoCal conceptual design consists of a highly granular and compact silicon–tungsten electromagnetic calorimeter supplemented by a spaghetti hadron calorimeter \cite{instruments6040070} covering the pseudorapidity region \mbox{$3.4<\eta<5.8$}.
The FoCal electromagnetic calorimeter is a sampling calorimeter segmented into 110 towers.
As illustrated in green in Fig.~\ref{detOverview}, one tower is made of 20 silicon pad (SiPad) sensors and has a length of 80\,m in the $z$-direction.
The basic building block of the electromagnetic calorimeter is the individual module (black rectangle in Fig.~\ref{detOverview}). 
Each individual module contains 5 towers built up from 20 alternating layers of passive absorber interleaved with active layers of silicon sensors (18 low-granularity with an individual pad size of $\sim 1\,\mathrm{cm^2}$ and two high-granularity CMOS Monolithic Active Pixel Sensor (MAPS) detectors with $\sim 30\times 30\,\mathrm{\mu m^2}$).

The pad layers provide a measurement of the shower energy and profile, while the pixel layers enable two-photon separation with high spatial precision (effective two-photon resolution of $d\sim 5\,\mathrm{mm}$) to discriminate between isolated photons and merged showers of decay photon pairs from neutral pions.

A fully instrumented tower prototype ($1/5$ of the final module size) was assembled to assess the detector performance for the chosen design and to address all foreseen technical issues in the final detector building such as Printed Circuit Board (PCB) production, sensor gluing, wire bonding, tungsten machining, mechanical assembly, etc.

The prototype tower is a sandwich of tungsten and silicon sensors covering a surface of \mbox{$9\times 8\,\mathrm{cm}^2$}.
The sensors are 320\,\textmu m-thick p-type silicon PIN photodiode arrays segmented into 72 cells (plus two calibration cells of smaller size) which were produced by Hamamatsu Photonics.

\section{Very front-end description}
\label{veryFEsSect}
As stated in the FoCal letter of intent \cite{focal_loi}, the best front-end solution for silicon sensor readout was identified to be the High Granularity Calorimeter Readout Chip (HGCROC) Application Specific Integrated Circuit (ASIC) developed for the CMS High Granularity Calorimeter (HGCal) \cite{CMS_upgrade,HGCROC_Thienpont_2020}.
This ASIC was designed to operate in the harsh radiation environment expected at CMS during the HL-LHC phase, much more stringent than the one expected at ALICE.
The HGCROC can instrument 72 regular cells and two special calibration cells, and monitor four common mode channels for coherent noise subtraction. 

Each channel has a low-noise preamplifier that converts the input charge from the silicon diode to a voltage signal.
This resulting signal is then sent to a shaper filter that feeds an Analog-to-Digital Converter (ADC) and to two discriminators connected to Time to Digital Converters (TDC).
The ADC path is used for signals up to 320\,fC, corresponding to $\approx 100$ Minimum Ionizing Particles (MIP) for a 320\,\textmu m-thick silicon sensor, and the TDCs are used by a time-over-threshold (TOT) measurement which provides the amplitude measurement for large signals and by a time-of-arrival (TOA) measurement which provides the timing information.
The combination of the ADC and TOT measurements allows for the coverage of a large dynamic range from 0.2\,fC to 10\,pC.
Furthermore, for calibration and testing purposes, an internal injection system can be enabled for each individual channel by slow control. 
The injection pulse is fired by a dedicated injection command through the fast command (FCMD) port.

The three measurements (ADC, TOA, and TOT) are sampled at the LHC frequency of 40\,MHz and stored continuously in a circular buffer (512 samples deep).
In parallel, the ADC and TOT measurements are used to compute digital sums (SUM) by groups of nine cells (resulting in eight sums per sample) for trigger generation.

A given entry of the circular buffer, featuring 72+2 measurements and four common modes measurements, can be transferred to the acquisition electronics by two serial links operating at 1.28\,Gbit/s.
This operation, which takes 1.075\,\textmu s, is requested by a readout command provided to the HGCROC through the FCMD port.
The entry to be transmitted is extracted from the circular buffer with an offset programmed by slow control.
The delay induced by the circular buffer, expressed in counts of machine clock cycles, has to be accounted for in order to read past values.
This delay is set in the initial configuration of the HGCROC.
The digital sums, also called SUM regions and compacted in two words of 32 bits at every bunch crossing, are continuously transferred by two serial links operating at 1.28\,Gbit/s.

The HGCROC uses a reference clock of 320\,MHz (an eight-time multiple of the LHC clock frequency), to build the required internal clocks for its operation and to synchronize itself to the LHC.
The fast commands (idle, readout, internal injection, synchronization) are transferred through an 8-bit serial link operated at 320\,Mbits/s and synchronous to the reference clock.
The configuration of the HGCROC 1210 internal 8-bit registers is achieved through an Inter-Integrated Circuit (I2C) link.

\section{Overview of the proposed electronics architecture for the full detector}
\begin{figure}[hbtp]
\centering
\includegraphics[angle=0,width=0.99\textwidth]{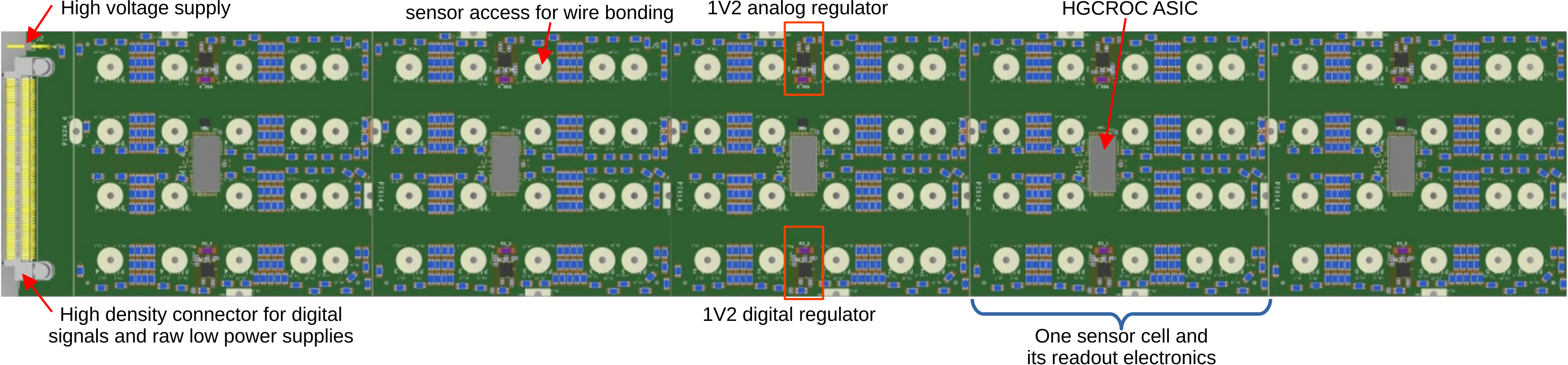}
\caption{
\label{fivePadFig} 
The 5-pad-layer board will feature one HGCROC per silicon pad sensor. Each HGCROC is accompanied by its own ultra-low dropout linear regulators to build its local power supplies.
Above each sensor cell, the access for the wire bonding is shown. 
The high power supply is provided by dedicated pins shown on the left side. 
The low raw power supply, is used to feed the linear regulators, and the various signals are connected to the high-density connector shown on the left side.
}
\end{figure}

Each pad layer of the FoCal electromagnetic calorimeter will be composed of 22 PCB having a dimension of about \mbox{$\rm 45 \times 8\,cm^2$}. 
Each PCB accommodates five silicon pad sensors glued on one side and has the very front-end electronic components mounted on the other side.
The silicon pad sensors are then wire-bonded to the 5-pad-layer boards through sensor access holes for wire bonding shown in Fig.~\ref{fivePadFig}.
Given the fact that these electronics will be exposed to radiation, and that the heat dissipation shall be minimized, the number of electronic components required to equip it should be kept as low as possible.

\begin{figure}[hbtp]
\centering
\includegraphics[angle=0,width=0.25\textwidth]{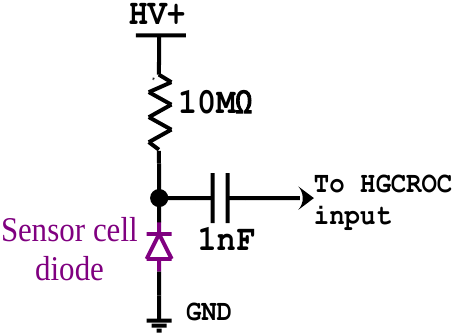}
\caption{
\label{sipad_biasFig} 
Schematic of one SiPad sensor cell bias.
}
\end{figure}

Hence, as shown in Fig.~\ref{fivePadFig}, the 5-pad-layer board is designed with one HGCROC per silicon pad sensor, each HGCROC has its own radiation-hard ultra-low dropout linear regulators to build its local power supplies and thus avoids any crosstalk between ASICs and mitigate any possible power supply noise conduction.
Each sensor cell diode is biased through a $\rm 10\,M \Omega$ resistor, and AC coupled through a 1\,nF capacitor with the HGCROC, see Fig.~\ref{sipad_biasFig}.
The high voltage can vary between 350\,V and 800\,V.

\begin{figure}[hbtp]
\centering
\includegraphics[angle=0,width=0.99\textwidth]{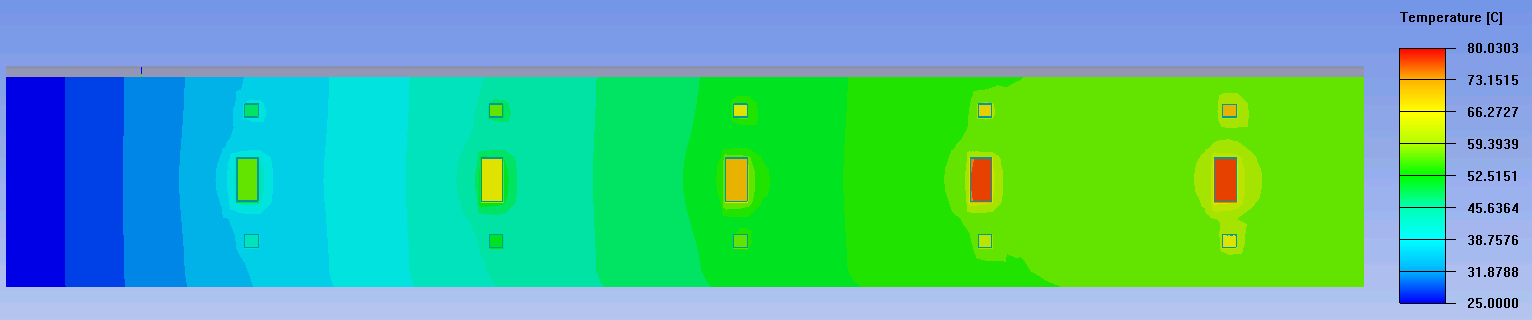}
\caption{
\label{thermalSimFig} 
Thermal simulation of a 5-pad-layer board (1.6\,mm thick) coupled to the silicon sensor (0.3\,mm thick) and the tungsten plate (3.5\,mm thick) with a total of 6.75\,W of heat injected on the main regions expected to generate heat (HGCROCs and their associated linear regulators).  
}
\end{figure}
With this architecture, the expected power dissipation is about 6.75\,W per board.
This power dissipation assumes a raw power supply of 1.5\,V, the usage of two linear regulators for supplying the locally regulated, and filtered 1.2\,V for analog and digital power supplies (600\,mA and 300\,mA respectively).
To assess the viability of this solution, a thermal simulation was performed by using the Ansys Icepak software \cite{AnsysSite} 
A simplified model of the 5-pad-layer board (1.6\,mm thick) coupled to the silicon sensor (0.3\,mm thick) and a tungsten plate (3.5\,mm thick) was used.
The power was injected on the main regions expected to generate heat (regulator surface of $\rm 5 \times 5\,mm^2$ and HGCROC BGA package surface of $\rm 17 \times 8\,mm^2$). 
The temperature elevation is predicted to be limited to 55\,\textcelsius\ with the limit condition fixed at a temperature of 25\,\textcelsius\ on the left edge of the board and with only conductive heat transfers considered.
The result of the thermal simulation is reported in Fig.~\ref{thermalSimFig}.
This estimated elevation is however an upper limit since the PCB was simulated with only a single 110\,\textmu m thick layer of copper sandwiched between two 800\,\textmu m thick layers of FR4 dielectric with no interconnecting vias. 

\begin{figure}[hbtp]
\centering
\includegraphics[angle=0,width=0.99\textwidth]{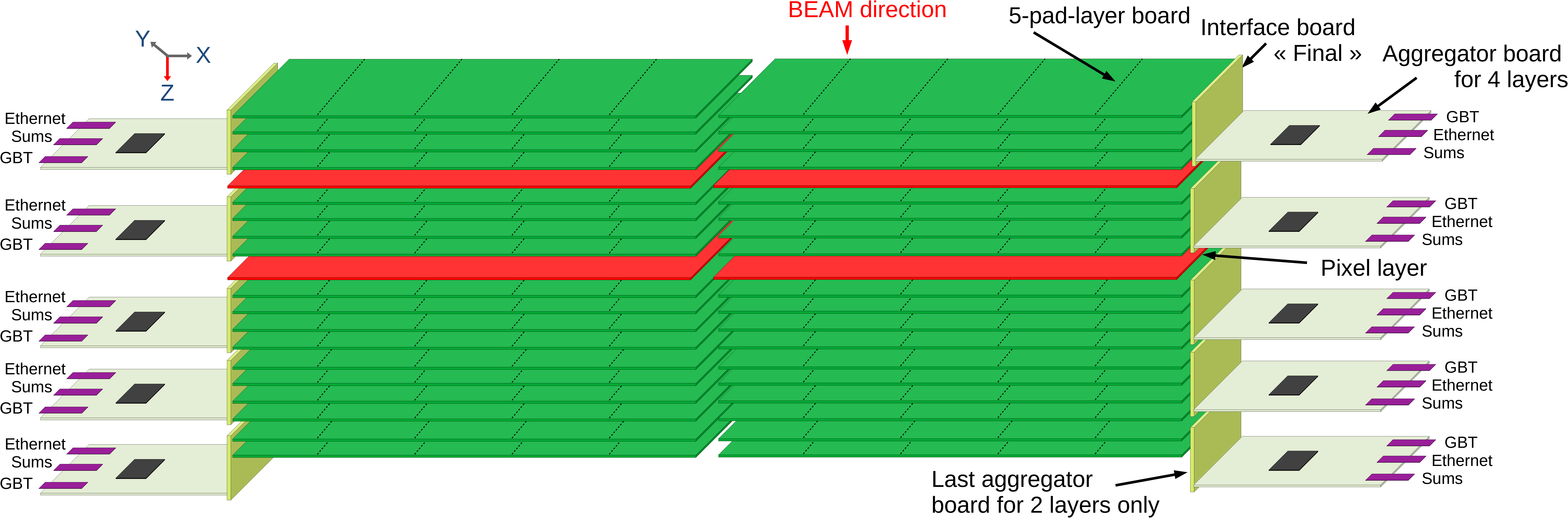}
\caption{
\label{stackFig} 
Overview of layer stacking in the beam direction.
The silicon pad sensor layers and their associated front-end electronics are shown in green, and the MAPS layers (without their readout) are shown in red.
The readout partition is shown for a full module: four 5-pad-layer boards per aggregator for the first four groups and two 5-pad-layer boards for the last silicon pad layers.
The mechanical adaptation between the 5-pad-layer boards and the aggregator is achieved with a dedicated `final' interface board.
}
\end{figure}

The 5-pad-layer boards will be read out on each side of the detector by FPGA-based aggregator boards.
Four 5-pad-layer boards will be connected to one aggregator through a dedicated `final' interface board as shown in Fig.~\ref{stackFig}. 
The aggregator board will be in charge of (i) controlling and reading out 20 HGCROC through Ethernet connection, (ii) establishing the bidirectional communication channel through the GigaBit Transceiver (GBT) protocol \cite{GBT_FPGA_paper} with the Common Readout Unit (CRU) \cite{CRU_paper}, (iii) sending the SUM information to a trigger board, and (iv) distributing the low and high voltage power supplies.
Several parameters were taken into account to determine the optimal readout partition of four layers per readout board.

Firstly, the average occupancy was estimated for each sensor layer as a function of the radial distance from the interaction point.
The most stringent conditions are met for each central 5-pad-layer board close to the beam pipe along the X axis, where the occupancy is estimated to be, respectively from the center to the edge, 12\%, 6\%, 1.5\%, 0.5\%, and 0.5\%.
Thus for a module of four 5-pad-layer boards, with 72 cells per sensor, 32 bits of data produced per cell, and an expected interaction rate of 1\,MHz in pp running during the LHC Run 4, the raw data rate is on average \mbox{$\rm 4 \times (0.12+0.06+0.015+0.005+0.005) \times 72 \times  32 \times 10^6=1.89\,Gb/s $} which fits the GBT capability of 3.2\,Gb/s\footnote{The data rate still fits the GBT capability even when taking into account the unusual word size of 80 bits when the Forward Error Correction (FEC) mode is used ($\rm  1.89\,Gb/s \times 80 / 64= 2.36\,Gb/s$)}.

Secondly, due to the insertion of MAPS on the fifth and tenth layers, it appears that the configuration with groups of four layers is the most convenient solution with regard to mechanical arrangement and reduction of the number of readout boards. 
A customized interface board would be needed only for the two last layers.

And lastly, a trade-off had to be found for the FPGA, between the input/output pin count, the amount of configurable logic, the availability of a reliable scrubbing functionality for radiation tolerance, and the component price.

Regarding a possible contribution of the FoCal to the ALICE trigger, it is envisioned to use only the SUM outputs of the second or third aggregator (located at mid-depth of the detector along the beam direction) where the electromagnetic shower energy deposition is expected to be the highest.
Therefore, using the information from only four silicon pad layers per module would be sufficient to produce an efficient trigger signal.
A dedicated SUM board in charge of collecting the refined information produced by the 22 aggregator boards would need to be designed to this end.
At this stage, it is foreseen to use active optical links between the aggregators and the SUM board. 

\begin{figure}[hbtp]
\centering
\includegraphics[angle=0,width=0.8\textwidth]{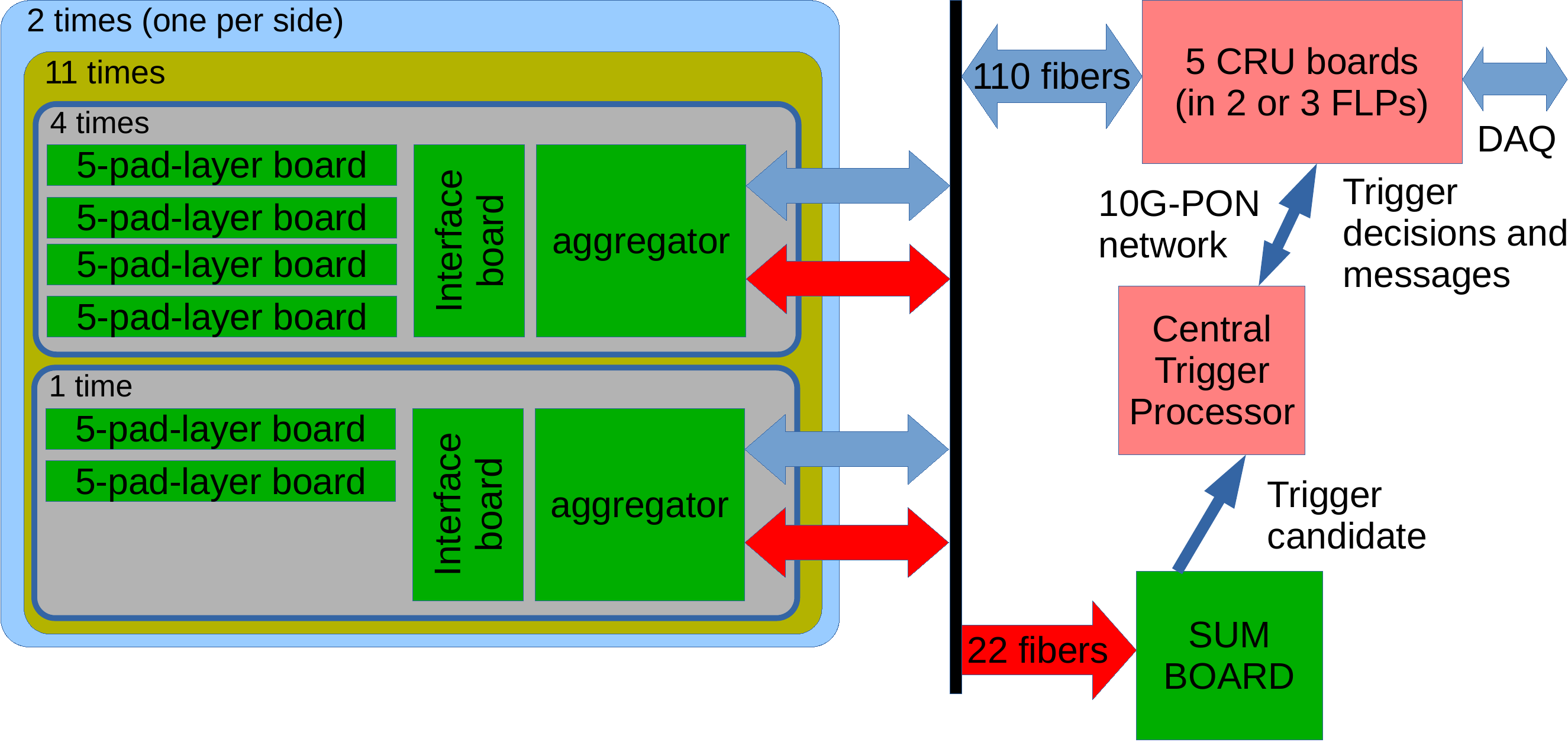}
\caption{
\label{electronicsArchFig} 
Overview of the electronics required to equip the full detector.
396 5-pad-layer boards, 110 aggregator boards, 110 `final' interface boards, and one SUM board are required.
Five CRU, set in two or three First Level Processors (FLP) are required to handle the full readout.
The electronics shown in green will be placed in the cavern.
}
\end{figure}

In total, 396 5-pad-layer boards, 110 aggregator boards, 110 `final' interface boards, and one SUM board will be required to equip the full detector (see Fig.~\ref{electronicsArchFig}). 
Given the fact that one CRU board can handle up to 24 GBT links, five CRU boards will be needed for the detector readout in the ALICE online system described in \cite{TDR_O2,Costa_2017}.

\section{Prototype electronics and associated tools}

\begin{figure}[hbtp]
\centering
\includegraphics[angle=0,width=0.8\textwidth]{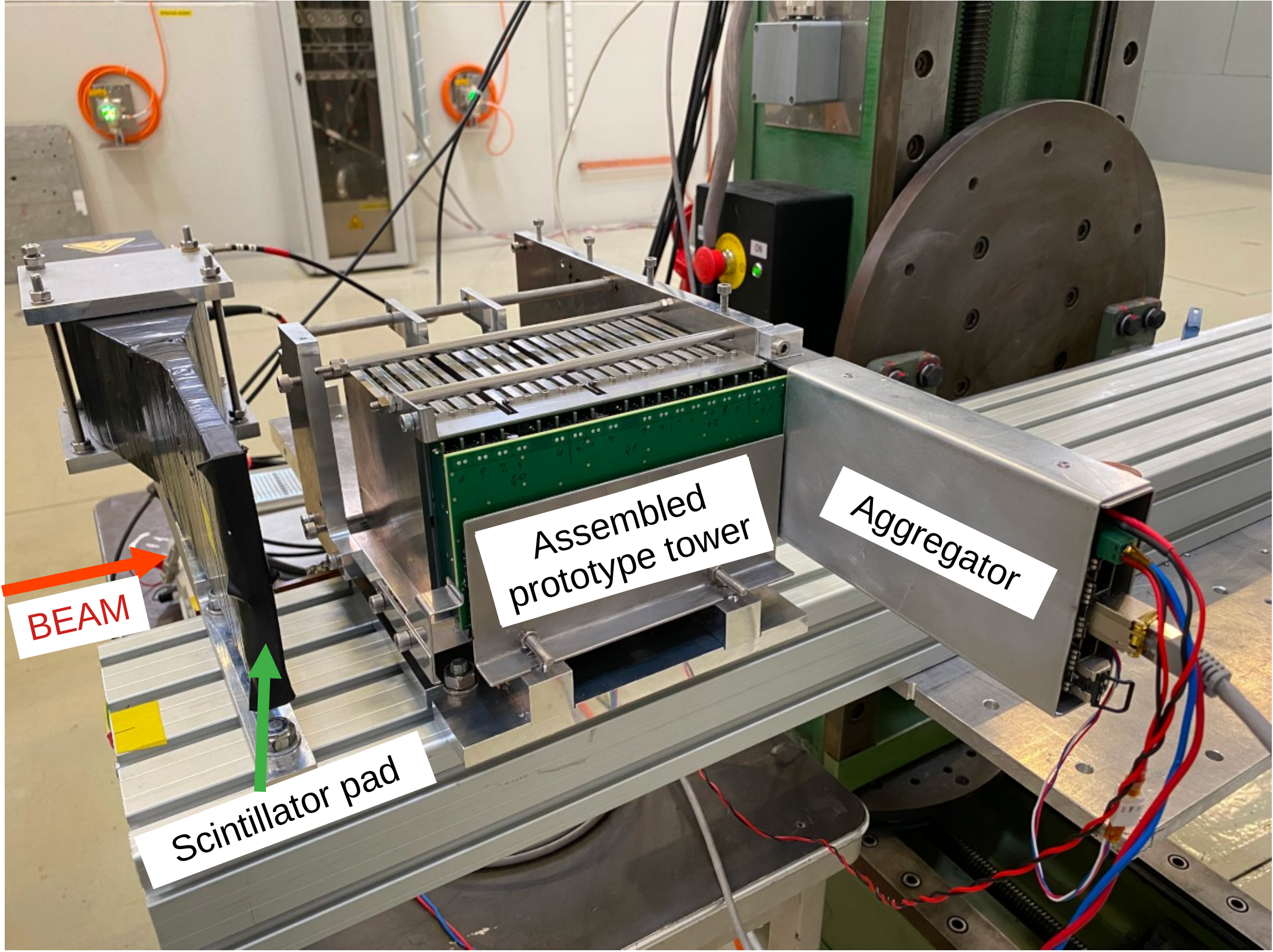}
\caption{
\label{protoTower} 
Picture of the FoCal tower prototype installed on the T9 beamline at the CERN PS.
}
\end{figure}

As illustrated in green in Fig.~\ref{detOverview} and shown in Fig.~\ref{protoTower}, a prototype tower ($\rm 90\,mm \times 80\,mm$) has been built including all 20 layers, 18 of which are silicon pad sensors.
Consequently, 18 single pad boards, a dedicated `prototype` interface board, and a prototype aggregator board were designed and manufactured.
The details about these developments and their associated test electronics are reported in the following sections.

\subsection{Single pad board design}
The single pad board corresponds to one-fifth of the final full-scale 5-pad-layer board.
Aside from its length, it was designed with the same constraints as those for the final board.
The single pad board allowed us to address the following critical aspects: height constraint, sensor coupling to the PCB, PCB flatness, high voltage handling, sensor grounding, and noise coupling mitigation. 
The description of the issues faced together with the adopted solutions is given hereafter.

\begin{figure}[hbtp]
\centering
\includegraphics[angle=0,width=0.75\textwidth]{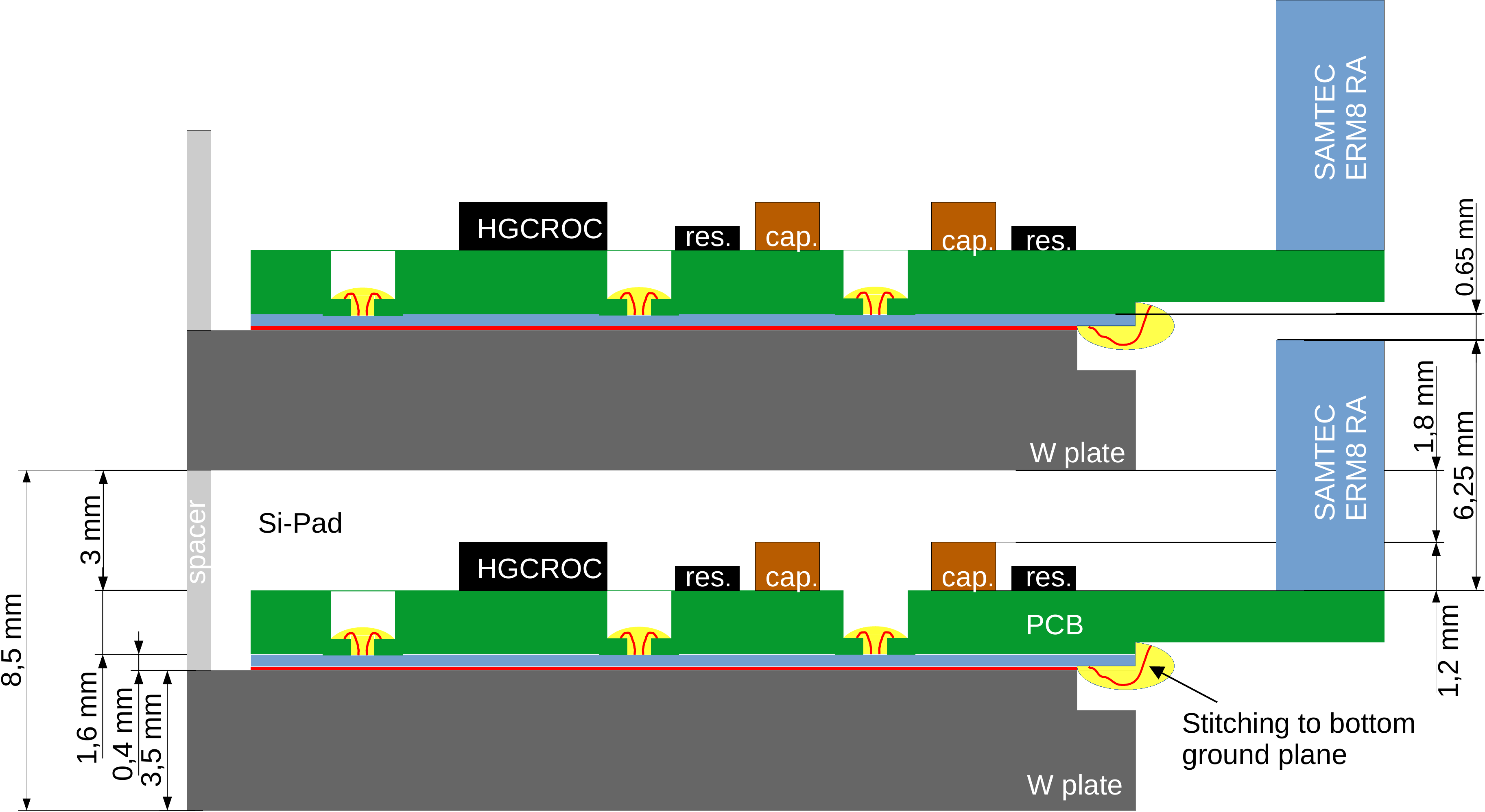}
\caption{
\label{sideViewFig} 
One pad layer cross-section view showing the constraint in height for the electronics and sensor stacking. The drawing is to scale in the height direction, but not in width.
}
\end{figure}

The single pad board uses components with a reduced height, in order to fit within the 5\,mm pitch of the sandwich made up of the PCB, the silicon pad sensor, and an adhesive sheet.
Additionally, a sufficient clearance with respect to the following tungsten plate must be kept to reduce the probability of high voltage arcing when operating the sensors at high bias voltages (about 500\,V).

Hence, the highest components on the board are the capacitors used for the capacitive coupling of the HGCROC to the Si-pad and for the high voltage decoupling (1206 form factor and 1.2\,mm in height).
The maximum thickness of the PCB was also constrained by the size of the high-density connector (6.25\,mm) and the pitch of 8.5\,mm between layers, as sketched out in Fig.~\ref{sideViewFig}.

\begin{figure}[hbtp]
\centering
\includegraphics[angle=0,width=0.7\textwidth]{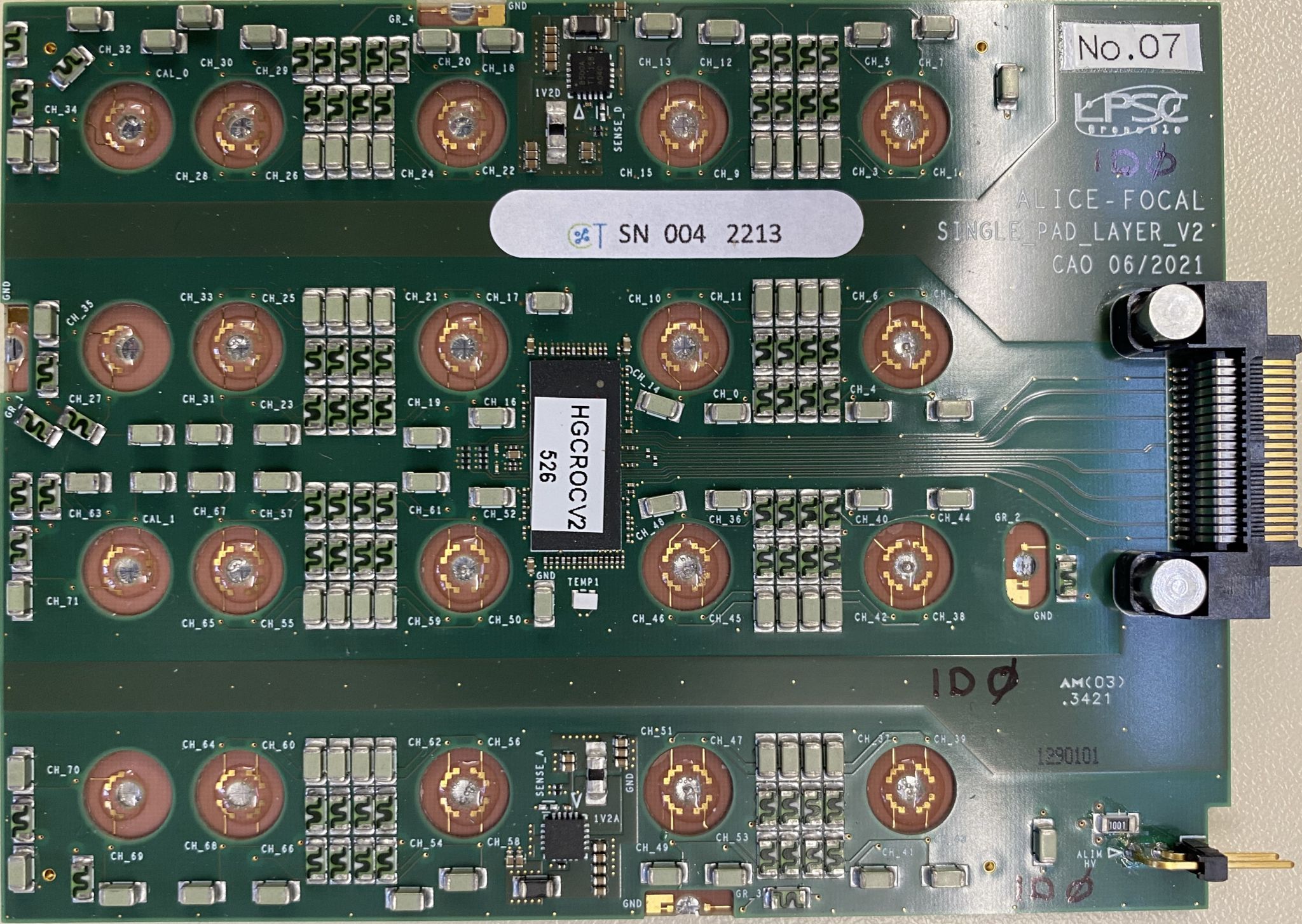}
\caption{
\label{singlePadPict} 
Single pad board designed with a total of 24 through holes for the sensor to PCB connection by wire bonding.
Each opening is composed of one hole with a shoulder, and the shoulder features wire-bond reception pads.
20 circular holes are used for the cell connection and four rectangular cavities located on the four sides are used for grounding and high-voltage connections.
}
\end{figure}

The single pad board was designed with a total of 24 through holes to allow the sensor to PCB connection by wire bonding.
Among these 24 cavities, 16 open up on four cells each, two on three cells, two on two cells, and four on guard rings for high voltage and ground connections.
To accommodate the wire bonder head access limitation (Devoltec G5 equipped with a deep access fine wire head 64000 DA), each cavity has been designed with a recess that features the wire-bond reception pads, see Fig.~\ref{singlePadPict}. 
For each cell, three reception pads per channel were implemented to minimize the parasitic inductance.
The passive component positioning with respect to the cavities was optimized to ensure sufficient access clearance for the bonding head.
The cavity diameter was reduced as much as possible to avoid warping caused by thermal retraction during fabrication.

\begin{figure}[hbtp]
\centering
\includegraphics[angle=0,width=0.99\textwidth]{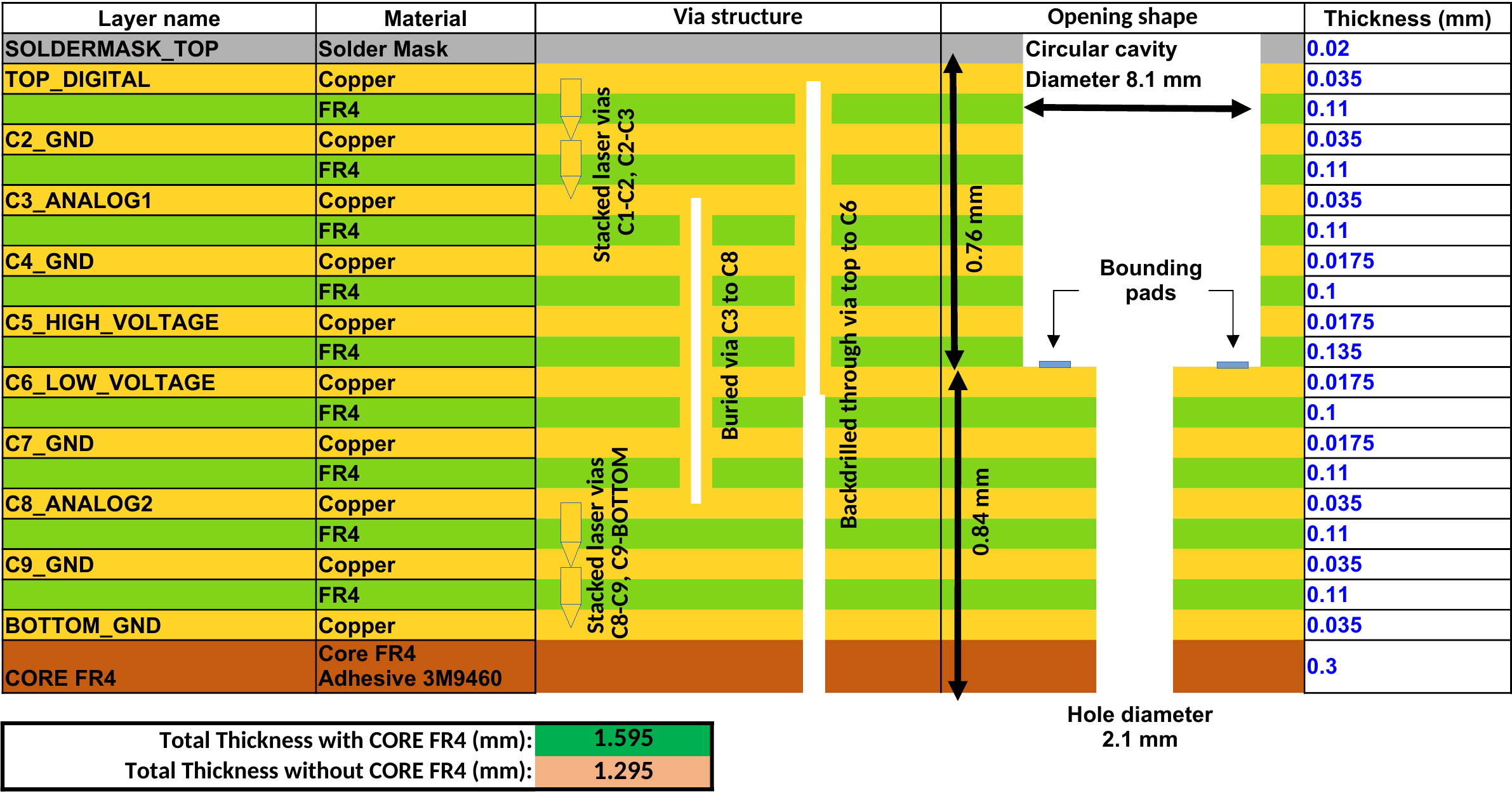}
\caption{
\label{layerStack} 
Cross-section of the single pad board displayed in Fig.~\ref{singlePadPict}. The materials of the alternating PCB layers with their respective thicknesses are shown. The transverse shape of a bonding hole is also represented in terms of layer extension and diameter sizes. 
}
\end{figure}
The single pad board PCB layer stacking, composed of ten copper layers, is shown in Fig.~\ref{layerStack} and described hereafter.
Five regularly interconnected layers are used as low-impedance ground.
Two layers are utilized for the HGCROC to silicon pad sensor connections and for routing the slow control and monitoring signals.
One layer is dedicated to the low-voltage power supply plane and to host the wire bond reception pads.
Another layer is used for the high-voltage plane. 
The top layer is used for the fast digital signals (frequencies in the GHz range).
Many ground planes are necessary to reduce cross-talk between the HGCROC channels and between digital and analog signals propagating in the different layers. 
Furthermore, copper symmetry must be ensured in order to avoid board warping during fabrication.
The via arrangement is selected to avoid any stubs for the high-speed analog signals between the wire bond reception pads and the top layer where the passive components and the HGCROC sit.
The last FR4 layer is used to insulate the sensor's high voltage from the bottom ground layer.
Thanks to the use of an FR4 adhesive core, it is pressed at ambient temperature and thus does not contribute to the board warping.
The thickness of this insulating layer should be as high as possible to minimize the parasitic capacitance seen by each pad cell.
The maximal thickness is however limited by the total PCB thickness which should allow the wire bonder head access to the bottom of the hole.
Therefore a thickness of 300\,\textmu m was used.

The boards were fabricated by two different manufacturers (in France and Japan), and it was observed that the deflection between the center and the edges was maintained below 10\,\textmu m on both productions.
This excellent result, well below the manufacturing tolerance (0.75\% of the PCB length), is to be compared to the deflection of about 1.5\,mm measured on the first PCB prototype.
In the previous version, rectangular-shaped and larger cavities were used to fit the passive components, and the copper layer stacking was asymmetric. 
On top of that, the first prototype suffered a fabrication flaw that led to the baking of the whole stacking (FR4 adhesive core included).

After gluing and wire bonding the sensor with the single pad layer board through the cavities, one side of the gold-plated face of the sensor (side in contact with the tungsten plate) is then stitched to the PCB bottom plane layer with a large number of wires.
The bottom side of the sensor is gold-plated in order to have a very good ground connection to the substrate. 
This minimizes both the resistance and the parasitic inductance return path for each cell.

In this prototyping phase, about 200 connections were used.
Even though the total number of wires could certainly be optimized, this step must be part of the sensor–board assembly process as it ensures a good sensor grounding necessary for its proper working.

The single pad boards were evaluated on a dedicated setup composed of an FPGA development kit, a test interface board to accommodate the interconnection between the FPGA Mezzanine Card (FMC) of the kit and the single pad board connector, and a multichannel charge injector.
These components of the setup are described in the following sections.

\subsection{Development kit firmware description}
\label{devKitFirmSect}
\subsubsection{Overview}
The development kit used was a KCU105 \cite{KCU105Guide} from XILINX\texttrademark{}. 
As a first step, this kit was used to assess the FPGA performance required for the FoCal readout.
In particular, it was necessary to confirm that regular inputs could be used to receive and deserialize the HGCROC data stream at 1280\,Mbit/s, and that the usage of the FPGA Mixed-Mode Clock Manager (MMCM) and its outputs were efficient to clock the HGCROC and send fast commands without degrading the HGCROC performance.
Another benefit brought by the usage of the development kit was the possibility to develop and validate the first critical aggregator firmware components before its manufacturing.

\begin{figure}[hbtp]
\centering
\includegraphics[angle=0,width=0.8\textwidth]{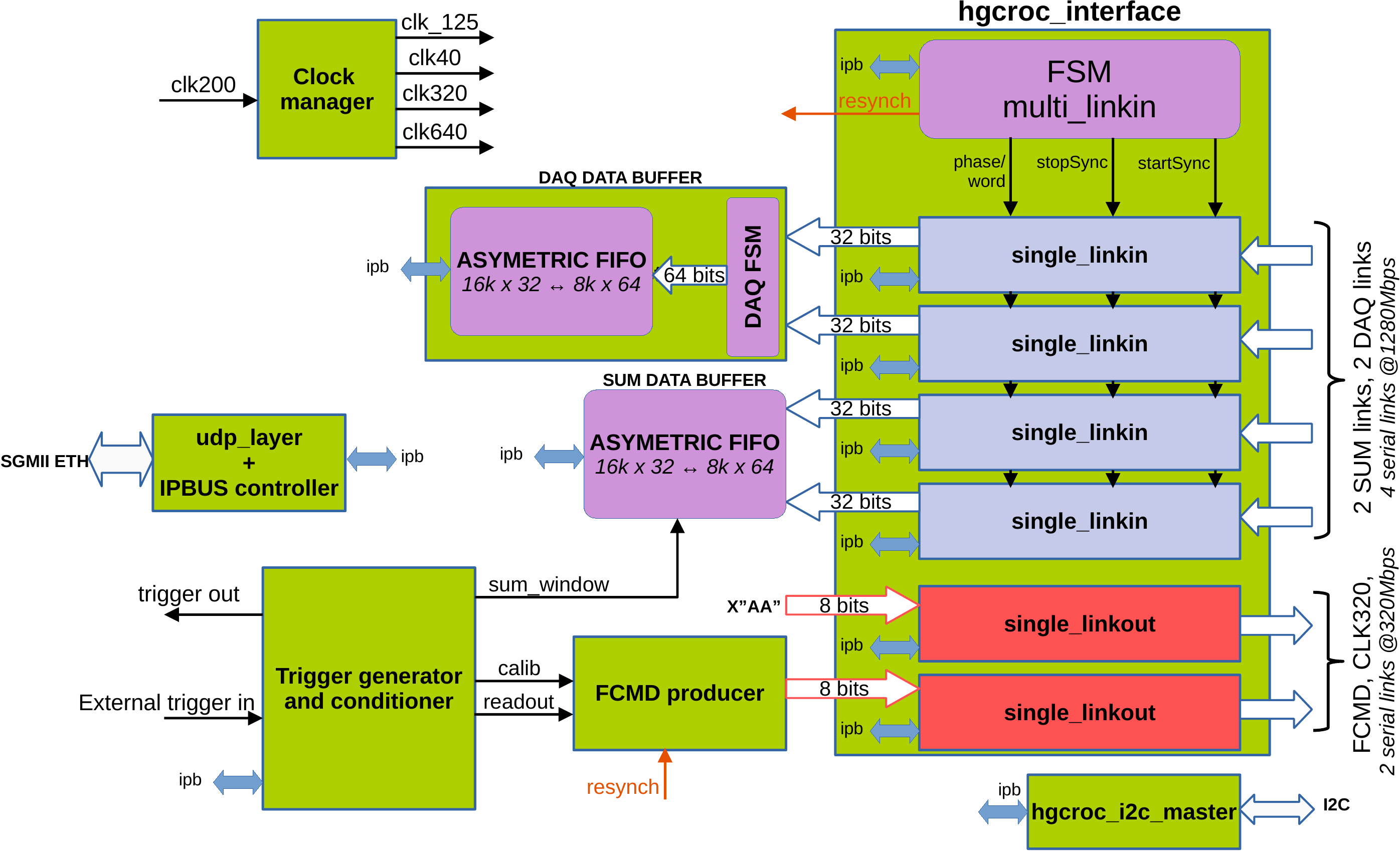}
\caption{
\label{devKitFig} 
Development kit firmware overview.
}
\end{figure}

A block diagram giving a structured overview of the firmware is provided in Fig.~\ref{devKitFig} and described in detail in the following. 
The clocks required for the HGCROC communication (40, 320, and 640\,MHz) and Ethernet operations (125\,MHz) are derived from a 200\,MHz local oscillator by utilizing two chained MMCM.

On the back-end side, the slow control and acquisition are achieved through IPBUS protocol \cite{IPBUS_publi} which provides a bridge between the physical layer of the Ethernet, through a Serial Gigabit Media-Independent Interface (SGMII), and an internal 32-bit address/32-bit data parallel bus, referred to as `ipb'.
This bridge is used to achieve the slow control of the firmware (parameter adjustment, synchronization management, I2C master control) and the acquisition of the two data buffers dedicated to the HGCROC SUM and DAQ links.

The HGCROC Inter-Integrated Circuit (I2C) master is designed to interface the IPBUS parallel interface and the I2C link of the HGCROC. 
The HGCROC requires three I2C access for every single register read or write access, and four for a read modify write access.
Consequently, the I2C master was designed to provide a higher protocol featuring the indirect addressing and access that can be triggered through a single IPBUS transaction.
Three high-level commands were hence implemented to speed up the configuration: read, write, and read modify write.

A dedicated block named `FCMD producer' is used to convert command requests into 8-bit words that are then serialized by the `hgcroc\_interface' and transferred to the HGCROC Fast Command (FCMD) port. 
This block manages four different commands: (i) the resynchronization command (resynch) which requests the HGCROC to send 255 times the synchronization word, selected by I2C, on all its output links, (ii) the calibration command (calib) which enables the internal charge injector (channels injected, injection capacitor, and charge value selected by slow control), (iii) the readout command (readout) that requires the HGCROC to send one full data frame on its DAQ links, and (iv) the idle command requiring no specific operation from the HGCROC. 

The `trigger generator and conditioner' can either be programmed to operate from an internal periodic trigger generator or to use the external input as a trigger source.
In both cases, a request to produce a readout command is sent to the fast command producer and, after an adjustable delay, the readout command request is issued.
After a different delay, a `sum\_window` signal, having a programmable duration, is produced and used as the write enable for the SUM data buffer. 
For test or HGCROC calibration purposes, it is possible to have a calibration command issued before the readout command and the SUM recording signal are generated.
All delays and durations are programmed in steps of 40\,MHz clock cycles.

The data produced by the HGCROC are stored in two asymmetric First In First Out buffers (FIFO).
The DAQ FIFO is managed by a Finite State Machine (FSM) which waits for a data header from the DAQ links to start the data frame recording and stores at 40\,MHz the two 32-bit words of data of the full data frame.
At the same time, the SUM link is continuously producing data, hence the recording is managed directly by the SUM recording signal produced by the `trigger generator and conditioner'.

\subsubsection{hgcroc\_interface description}
The `hgcroc\_interface' is in charge of providing the HGCROC with its reference clock, sending the Fast Command stream, and receiving and deserializing the SUM and DATA streams.
This block is generic by design and for this reason, it uses elementary cells named `single\_linkin' and `single\_linkout' to perform respectively (i) the deserialization from a serial stream of 1280\,Mbit/s to 32 bits clocked at 40\,MHz and (ii) the serialization from 8 bits clocked at 40\,MHz to a serial stream of 320\,Mbit/s.
The `single\_linkin' blocks are all simultaneously controlled by a common FSM which handles each sequence of the HGCROC link synchronization procedure.
Each sequence starts by setting the HGCROC in synchronization mode with a dedicated resynchronization command (resynch) decoded by the HGCROC. 
A synchronization word is then transmitted 255 times by the HGCROC to the FPGA inputs.
The sequence ends by checking the proper reception of 255 words.

\begin{figure}[hbtp]
\centering
\includegraphics[angle=0,width=0.9\textwidth]{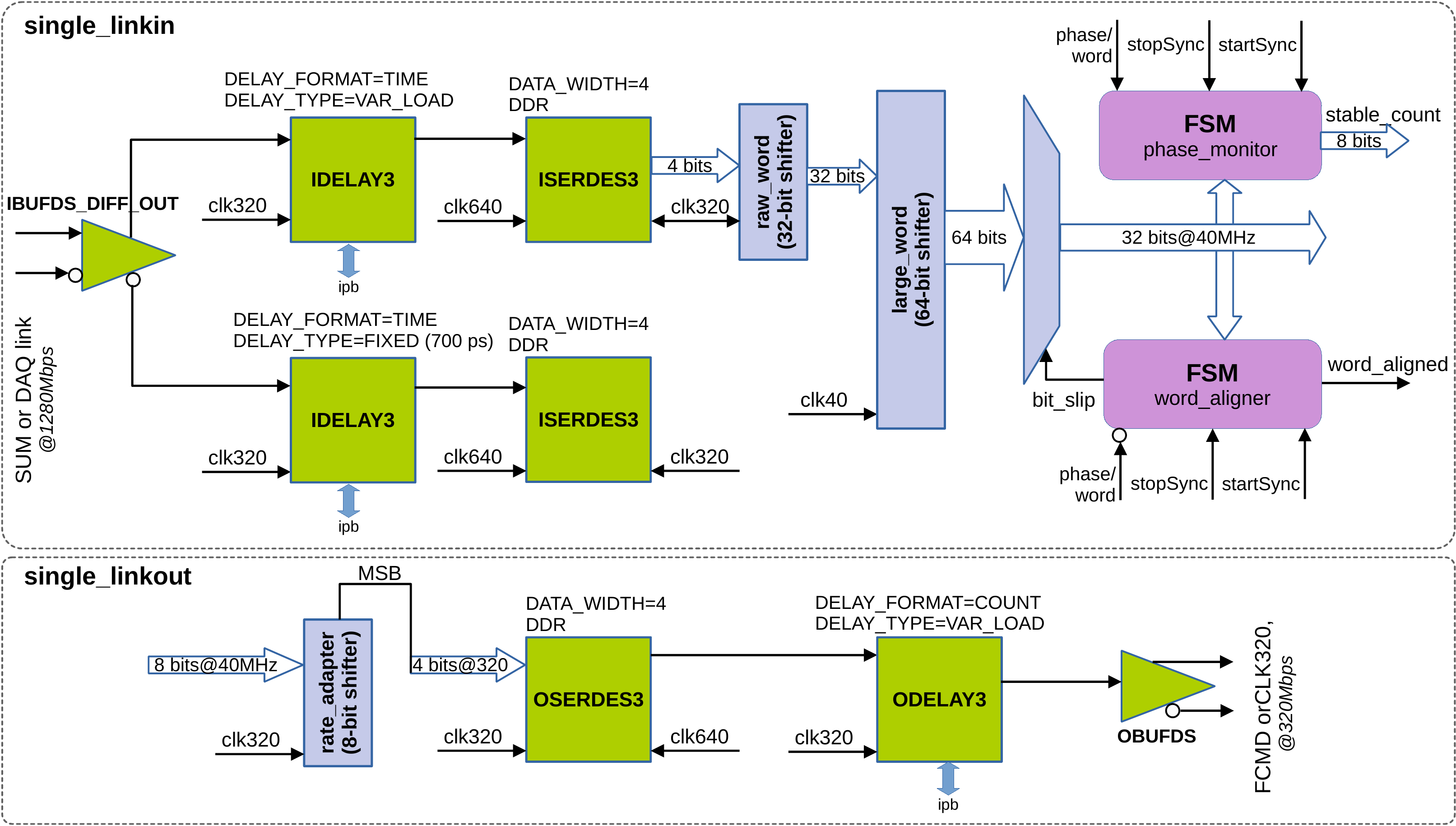}
\caption{
\label{singleLinksFig} 
Overview of the `single\_linkin' (top) and `single\_linkout' (bottom) blocks implementation.
In green, the resources provided by the XILINX\texttrademark{} Ultrascale FPGA \cite{selectioGuide}, in light blue the shift registers and the 32-bit selector, and in magenta, the FSMs dedicated to link synchronization.
}
\end{figure}
Figure~\ref{singleLinksFig} gives more details about the implementation of the `single\_linkin' and `single\_linkout' blocks.
The serial link from the HGCROC enters the `single\_linkin' through a differential input buffer that feeds two input delay lines (ISERDES3) followed by 4-bit dual data rate deserializers (ISERDES3) operated at 640\,MHz.
This structure was chosen to permit operation in the so-called `time mode' which requires delay line calibration as recommended in \cite{selectioGuide}.
With this structure, the tap value required in the IDELAY3 cell to obtain a 700\,ps delay is read via the IPBUS interface.
This measured value is then used to compute the delay per tap by the driving software, which uses it as input for the link synchronization procedure.
The 4-bit deserialized word is then transferred to a shift register (`raw\_word') that shifts its previous content by four positions and loads the new four bits at 320\,MHz.
The `raw\_word' output is sampled by a 64-bit shift register that shifts its previous content by 32 positions and then loads the 32 new bits at 40\,MHz.
Finally, a 32-bit portion of this last register is selected by a multiplexer controlled by the `word\_aligner' FSM.
Two FSMs, controlled by the 'multi\_link' FSM, are used to monitor the 32-bit output. 
The `phase\_monitor' FSM counts the number of equal word receptions, which should reach the maximum value of 255 if the input serial link is properly sampled during the re-synchronization sequence. 
In a second step, after the delay has been properly adjusted, the `word\_aligner' searches incrementally among the 32 possible values for the proper `bit\_slip' value, and once found, it flags a success with the `word\_aligned' port.

The `single\_linkout' block takes the 8-bit word provided by the `FCMD producer' and serializes it on the 40\,MHz clock, loads it and  shifts eight times at 320\,MHz.
The MSB of the shift register (`rate\_adapter') is then serialized four times at 1280\,MHz, which results in the 320\,Mbit/s rate that is transmitted through an output differential buffer (OBUFDS).
This complexity is due to the fact that the usage rules of the FPGA imposed the same frequencies on all delay blocks, and besides, the SERDES block had to have the same divided clock frequencies as their associated DELAY blocks.

\subsection{Charge injector board}
The silicon sensor assembly with the single pad board is done by wire bonding which is a non-reversible operation.
Therefore, a charge injector board was developed to test and characterize the front-end boards before wire bonding the sensors.
In the longer term, the same charge injector board could be routinely used for testing the PCB production and component assembly.
The charge injector board was designed to be tied to an unbonded single-pad board with a spring-loaded contact (mill-max\textsuperscript{\textregistered} \mbox{0906-0-15-20-76-14-11-0}).
\mbox{72 + 2} of these were used for charge injection and four were used for ground connection on the periphery. 

\begin{figure}[hbtp]
\centering
\includegraphics[angle=0,width=0.45\textwidth]{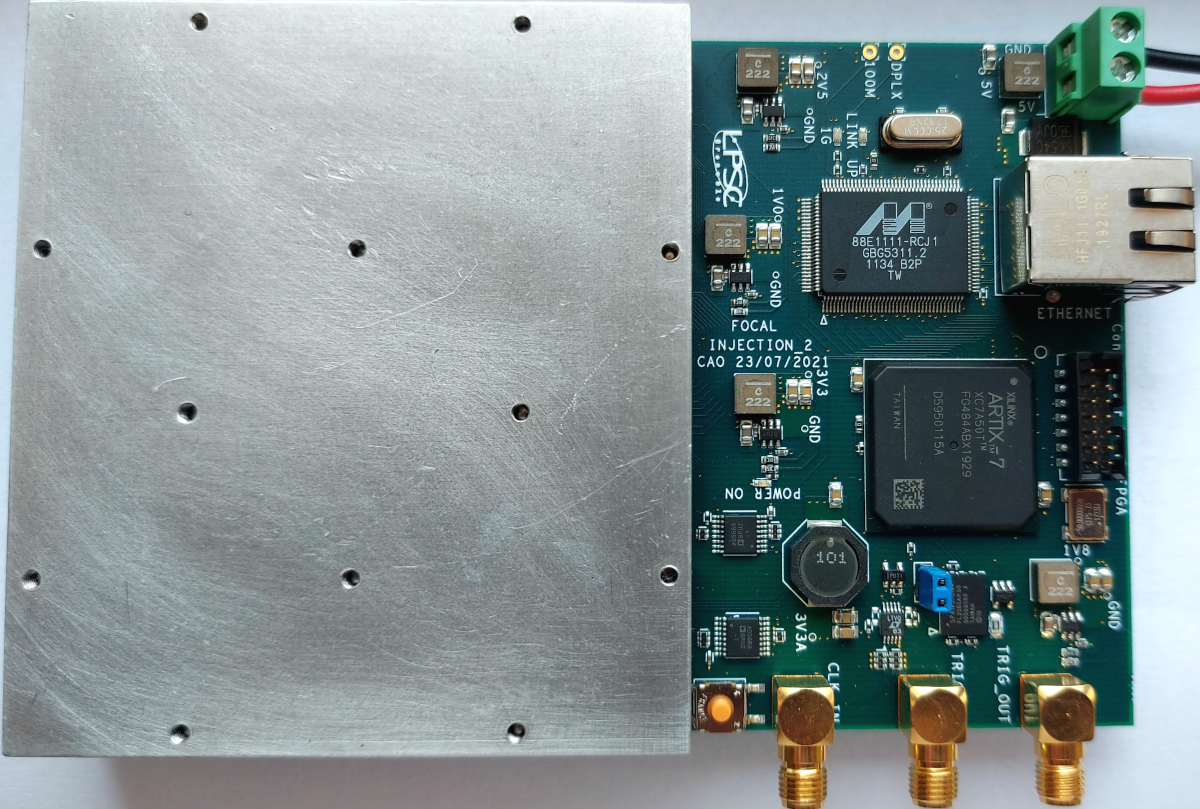}
\includegraphics[angle=0,width=0.49\textwidth]{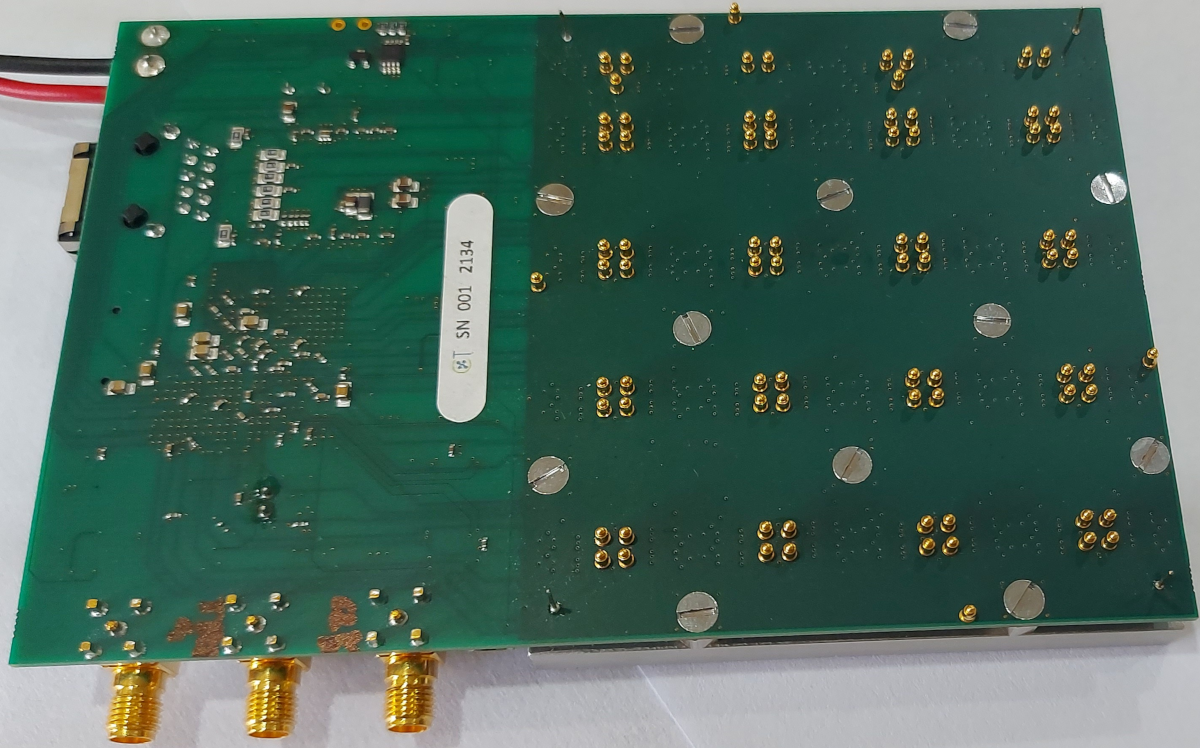}
\caption{
\label{injPic} 
Top and bottom views of the charge injector board shown respectively on the left and right-hand sides.
As shown on the left, a metal plate was added to stiffen the injector board and avoid deformation when pressing the injector on the single pad board.
}
\end{figure}

The board consists of 72\,+\,2 charge injectors individually controlled by an FPGA (XILINX\texttrademark{} \mbox{XC7A35T-2FGG484C}) and trimmed by Digital to Analog Converters (DAC), see Fig.~\ref{injPic}.
The FPGA provides an IPBUS bridge to ensure the configuration of the firmware (channels to fire, pulse duration) and of the DACs through serial peripheral interface protocol.
To ensure a synchronous operation with the HGCROC, the board features a 40\,MHz clock input, and trigger input, both equipped with DAC adjustable threshold comparators.
A total of eleven outputs of two octal 16-bit DAC are used to adjust the channel injector setpoint values and two others are used to adjust the clock and the trigger input comparator thresholds.
The channel setpoint values are distributed as follows: nine DAC outputs to adjust a one-cell channel in each SUM region (see section~\ref{veryFEsSect}) allowing for different levels of injection within a single region (eight in total) and the two remaining are dedicated to the calibration cell channels.

\begin{figure}[hbtp]
\centering
\includegraphics[angle=0,width=0.45\textwidth]{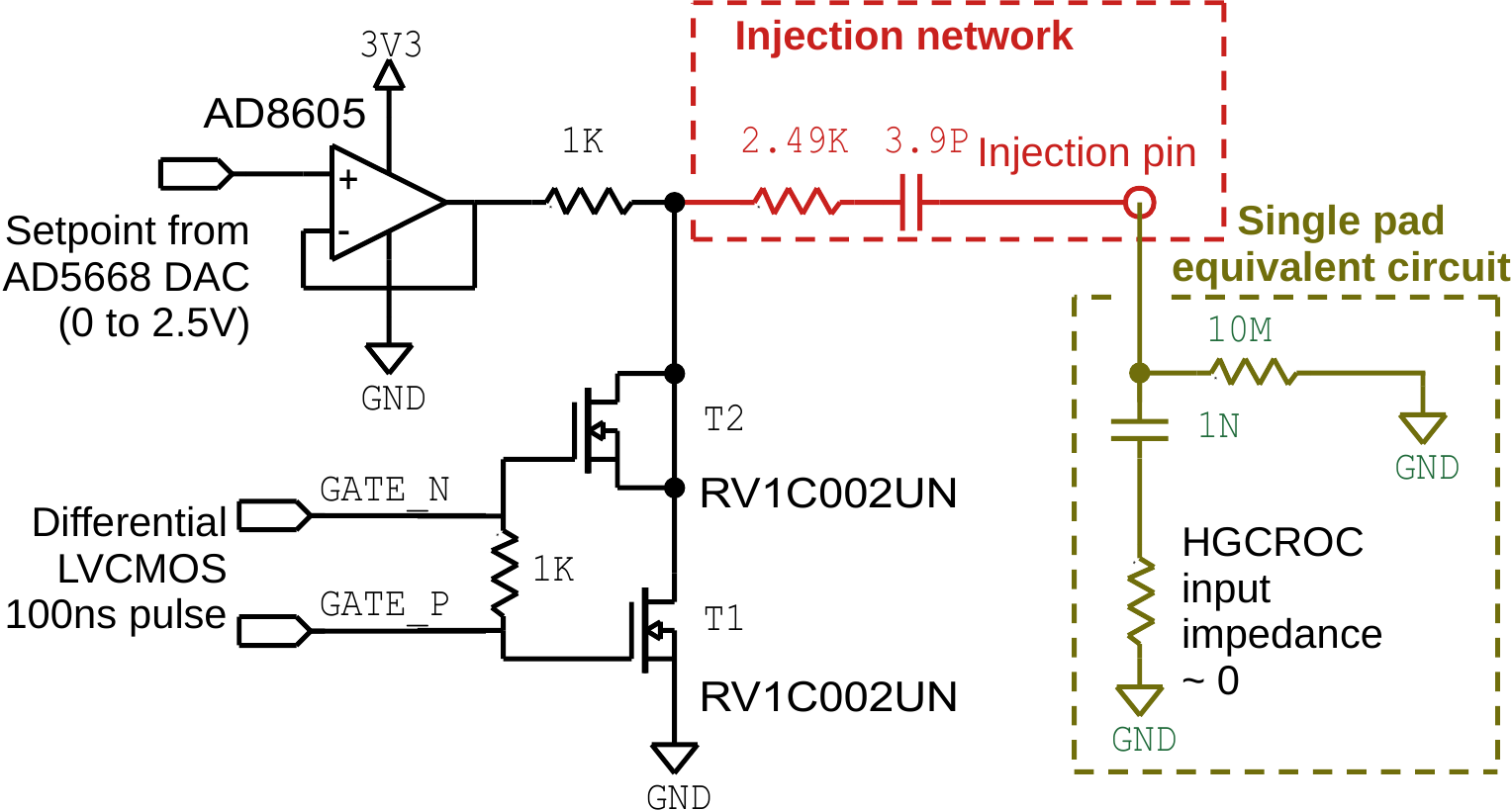}
\includegraphics[angle=0,width=0.54\textwidth]{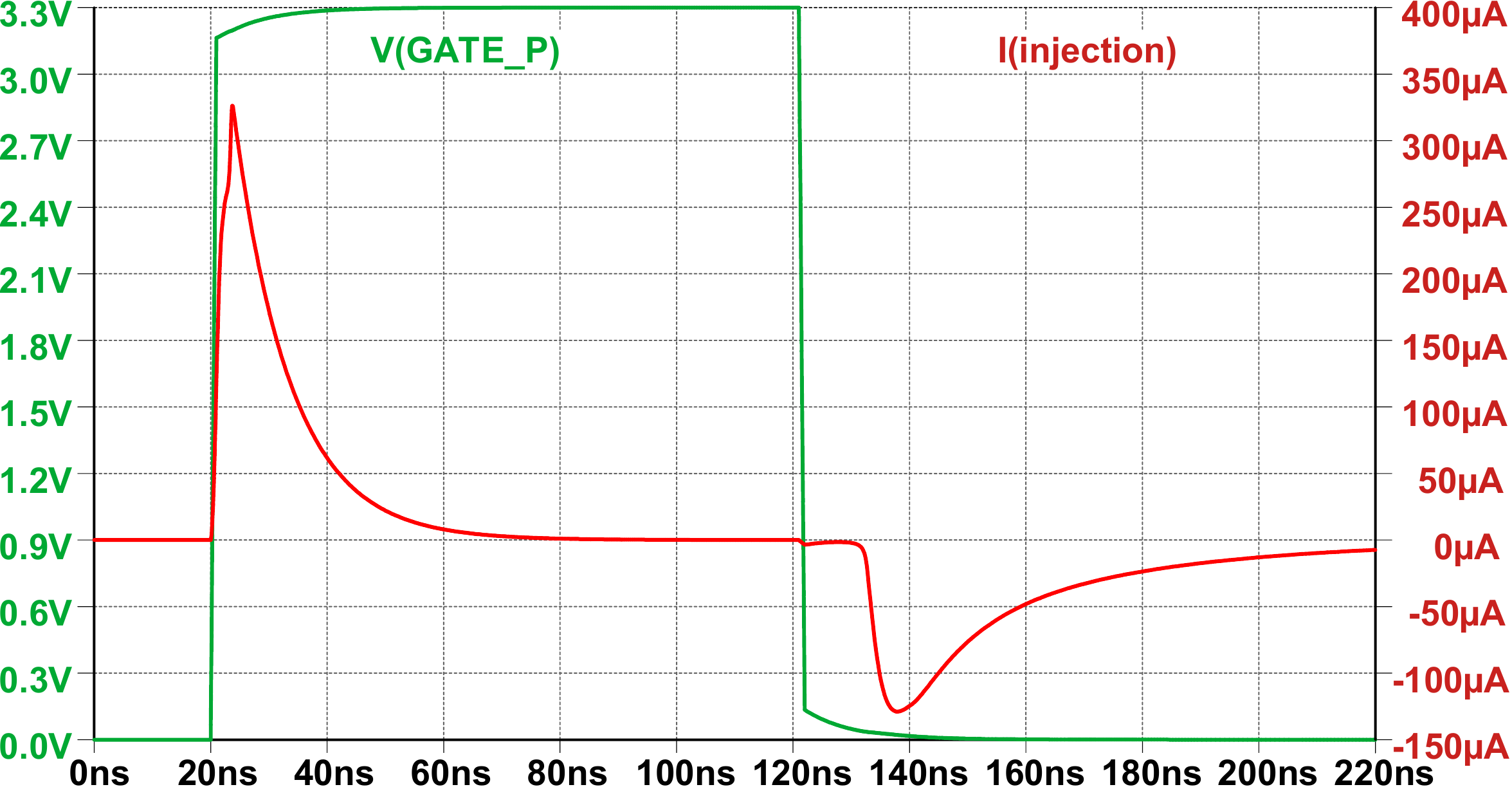}
\caption{
\label{injAnalogFig} 
Schematic of a single channel charge injector on the left: the injection network is shown in red, and the equivalent model of the HGCROC input and the single pad board is shown in dark green.
SPICE simulation of the channel injector on the right: the positive GATE signal is shown in green, and the injected current is in red with a DAC setpoint of 1\,V.
}
\end{figure}
The schematic of a single-channel injector circuit is shown on the left-hand side of Fig.~\ref{injAnalogFig}.
The principle of operation is the following, a small capacitor (3.9\,pF) is loaded to a given voltage, thanks to the 16-bit DAC setpoint value (0 to 2.5\,V).
Hence, the stored charge is equal to capacitance (3.9\,pF) times the loading voltage, which yields a maximum stored charge of about 10\,pC.
To inject a charge,  the injection capacitor is grounded through a resistor meant to limit the current pulse amplitude and thus stay within the operational requirement of the HGCROC.
To achieve that, the MOS transistor T1 is driven with a Low Voltage CMOS (LVCMOS) level for a duration of 100\,ns.
The transistors ensuring the grounding during injection were selected to be of very low input capacitance (typically 12\,pF) to minimize parasitic charge injection from the grid.
In addition, in order to cancel the parasitic charge injected through Grid-Source and Grid-Drain capacitors of transistor T1, a second transistor (T2) is driven with a complementary signal.
The matching of T1 and T2 parasitic capacitance is critical to perfectly compensate for the charge injection. 
As this is difficult to achieve, low-input capacitance transistors were chosen.
A SPICE simulation result of this circuit is shown on the right of Fig.~\ref{injAnalogFig}.

\begin{figure}[hbtp]
\centering
\includegraphics[angle=0,width=0.59\textwidth]{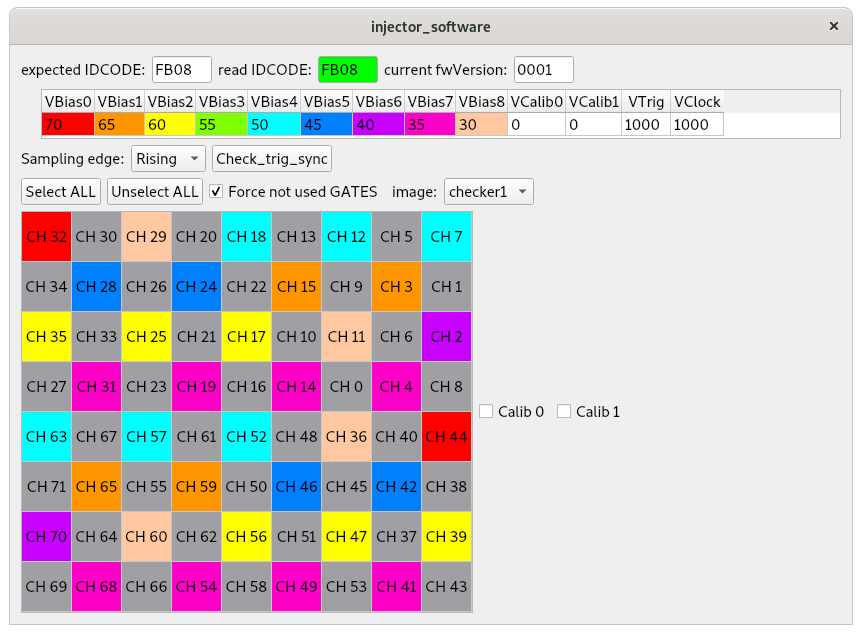}
\includegraphics[angle=0,width=0.4\textwidth]{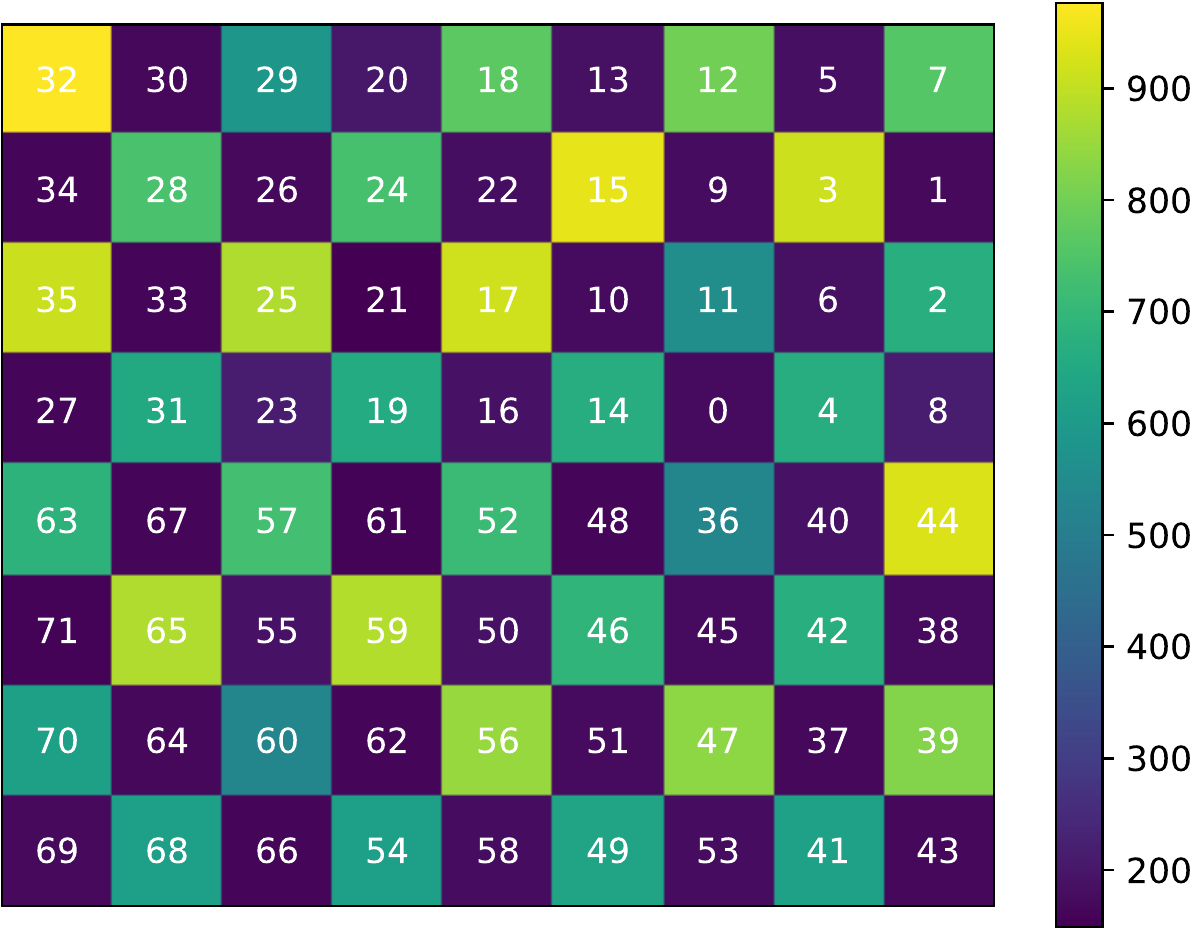}
\caption{
\label{injCheckerFig} 
Screenshot of the injector driver software Graphical User Interface (left) configured to inject a checker pattern with different charge injection levels.
Deactivated channels are grayed, and activated channels are colored in accordance with the bias voltage applied.
On the right, the average value recorded on each channel is shown in a two-dimensional plot. 
The matching can be observed.
}
\end{figure}

Dedicated Qt-based software was developed to drive and set up the injection board, see Fig.~\ref{injCheckerFig}.
The software allows the setting of each bias voltage in mV and the individual selection of the channels to be injected.
Consequently, on the left of Fig.~\ref{injCheckerFig}, deactivated channels are grayed and activated channels are colored in accordance with the bias voltage applied.
Each single pad board was tested before assembly with four injection patterns (no injection, checker pattern, complementary checker pattern, and all channels).
The recorded data were confronted to the injection patterns to ensure proper successful operation, an example is shown on the right of Fig.~\ref{injCheckerFig} for the checker pattern.

\subsection{Aggregator board}
\subsubsection{Hardware description}
In the final system, the aggregator board will be in charge of controlling and reading out 20 HGCROCs (four 5-pad-layers boards) of the FoCal pad detector and managing the back-end communication with the ALICE Central Trigger Processor (CTP) \cite{CTP_Krivda_2016} and O2 data acquisition system \cite{TDR_O2,Costa_2017}.
Given the fact that the tower prototype has about the same number of HGCROCs to manage (18), the board was designed to be fully compatible with the two layouts, the interface board format being the only difference: i.e. either interfacing one aggregator with 18 single-pad boards arranged in a tower or one aggregator with four 5-pad-layer boards arranged in the final FoCal module.

\begin{figure}[hbtp]
\centering
\includegraphics[angle=0,width=0.95\textwidth]{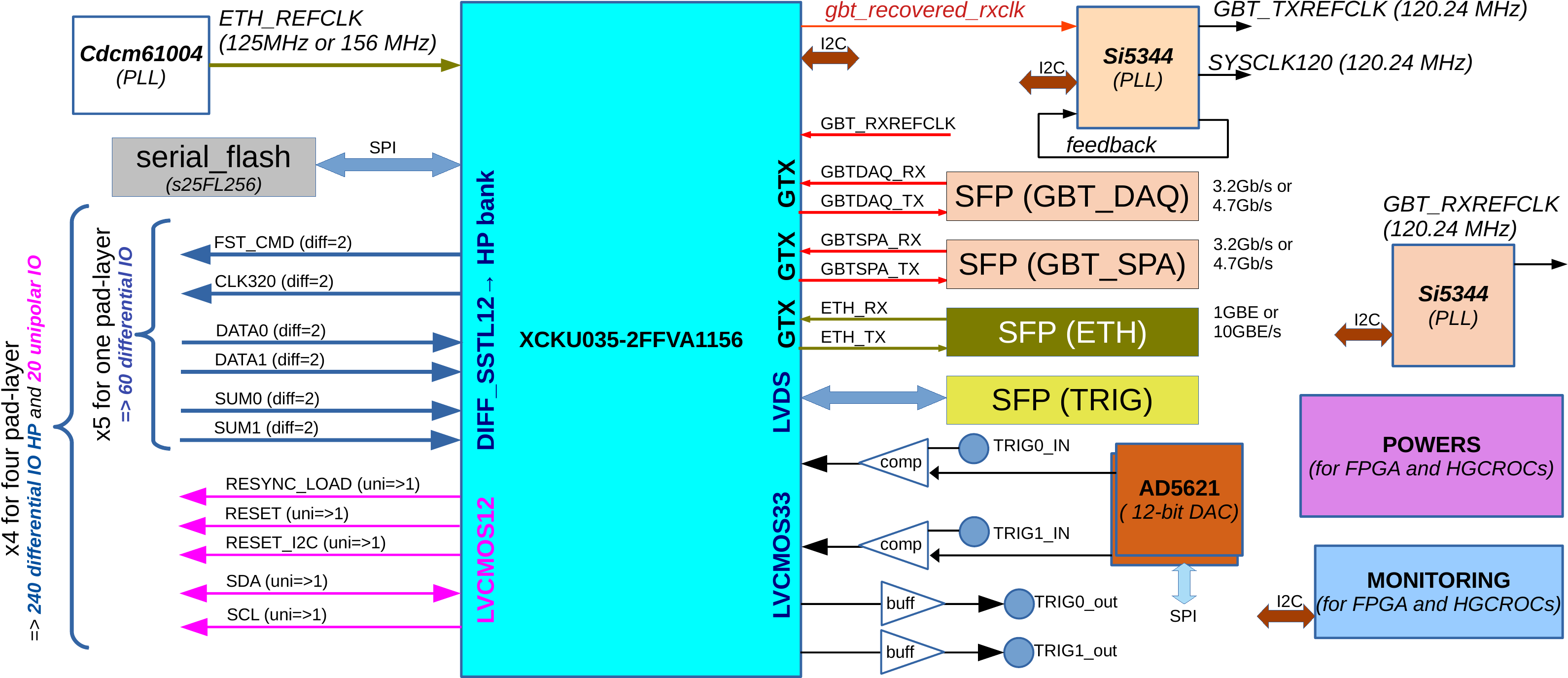}
\includegraphics[angle=0,width=0.9\textwidth]{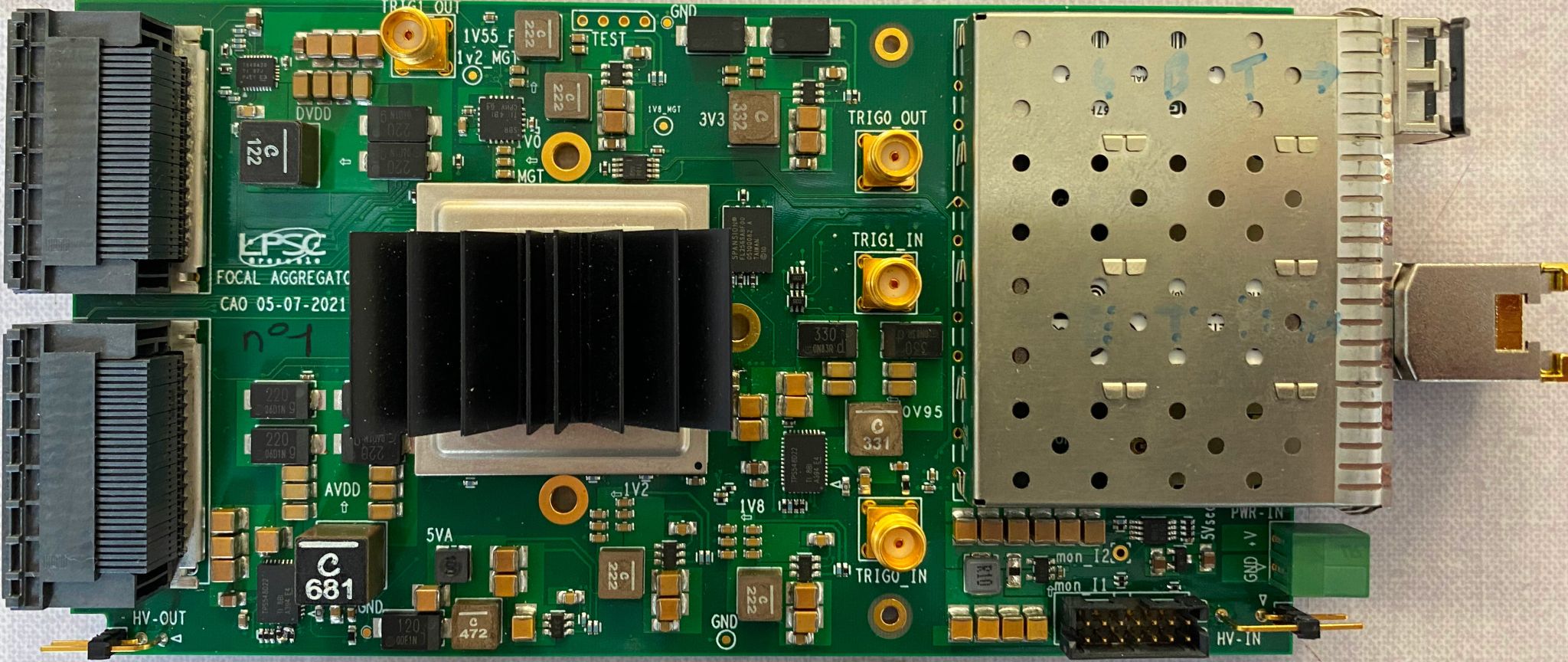}
\caption{
\label{aggregatorHardFig} 
Hardware overview (top) of the aggregator board with the main interfaces displayed and picture of the fabricated board (bottom).
The PCB has a dimension of $\rm 80\,mm \times 150\,mm$, accounting for the Ethernet module, and the length extends to 180\,mm.
}
\end{figure}
Figure~\ref{aggregatorHardFig} shows a block diagram of the aggregator board hardware parts with the main interfaces displayed (top) and a picture of the fabricated board (bottom). 
The core component of the board is an FPGA from XILINX\texttrademark{} \mbox{XCKU035-2FFVA1156}.

On the front-end side, the FPGA handles a total of 120 differential high-speed serial links in differential 1.2\,V Stub Series Terminated Logic (SSTL) level and 20 slow control links in 1.2\,V Low Voltage CMOS (LVCMOS) level. 
The unipolar signals are shared by one 5-pad-layer board (embedding five HGCROCs) or five single-pad boards.
The differential signals are carefully routed point-to-point with a controlled differential impedance of 100\,$\Omega$.

On the back-end side, the FPGA is connected to four Small Form-factor Pluggable (SFP) modules hosted in a quad slot cage.
Two of those SFP (one nominal, one spare) modules are dedicated to optical transceivers for the GBT link, one is for the Ethernet transceiver (copper or optical), and the last one is for communication with the SUM board through an Active Optical Cable (AOC).

Additionally, the aggregator board provides two discriminated inputs with adjustable thresholds used for external triggering plus two digital outputs that can be used to fire a light injector (Infra Red LED) or drive the charge injector (clock and trigger).

Concerning the support functionalities, the FPGA is equipped with a Serial Peripheral Interface (SPI) flash memory that holds its firmware and the Ethernet identification information (MAC number).
The serial flash can be updated remotely through the Ethernet connection. 
The board features various power converters to locally generate the various power supplies required for the aggregator and the 5-pad-layer boards from the main 12\,V.
The aggregator also conveys the high voltage (350\,V to 500\,V) to the front-end boards.

The clocking is separated into two main domains, one for the system configuration and communication through Ethernet and the other dedicated to the acquisition clock domain.
The low-jitter, high-precision Phase Locked Loop (PLL) components (SI5344) are configured via I2C by the FPGA which receives their configuration by slow control.
The Ethernet reference clock is by default set to 125\,MHz to permit 1\,Gb Ethernet (GbE) but can be increased to 156.25\,MHz to permit 10\,GbE.
This increase is possible provided the firmware is adapted and an appropriate SFP transceiver module is selected.
This reference clock is a free-running clock that is only hardware configured to ensure continuous operation and Ethernet communication.
The acquisition clock domain is based on either an onboard generated 120.24\,MHz clock (three times the LHC reference clock frequency of 40.08\,MHz) or on the GBT link recovered clock, depending on the firmware configuration.
The selected acquisition clock is used for the FPGA acquisition part and to clock the HGCROCs.

The aggregator also features an extended monitoring functionality. 
The FPGA itself can monitor its own temperature and voltage through the integrated system monitoring feature \cite{sysmonGuide}.
The board is equipped with ten additional 8-channel 16-bit sigma-delta ADCs (ADS114S08B) for deep monitoring of the four 5-pad-layer operating parameters.
This results in the monitoring of 20 voltage and 20 current values and three values of temperature per layer (12 in total).

The total power draw of the prototype, with 18 single-pad boards connected and configured, and with the firmware described in section~\ref{aggFirmSect}, was measured to be about 36\,W.
Subtracting the power consumed by the single pad boards of about 24.3\,W, the aggregator board power consumption itself was estimated to be approximately 12\,W.

\subsubsection{Firmware description}
\label{aggFirmSect}

\begin{figure}[hbtp]
\centering
\includegraphics[angle=0,width=0.99\textwidth]{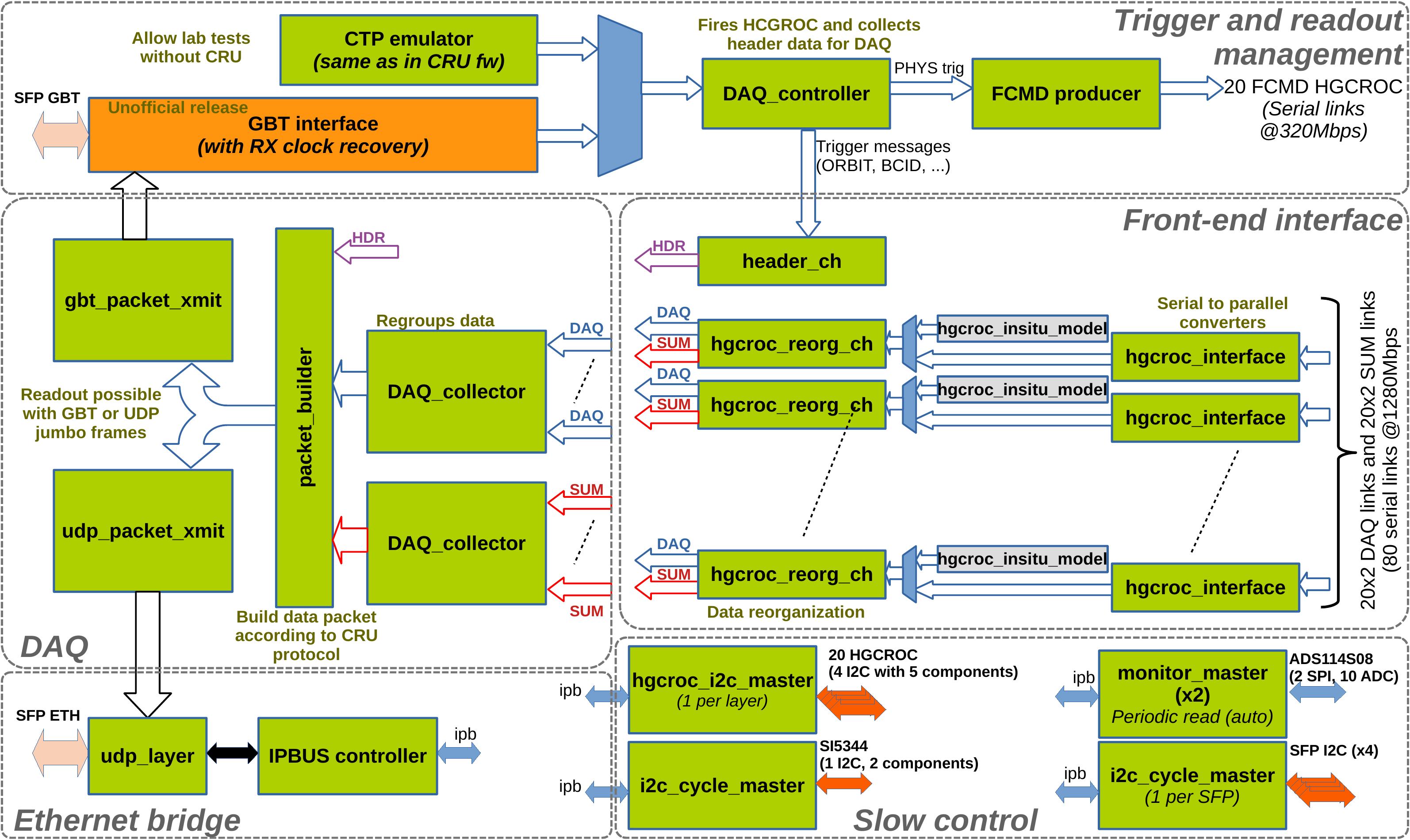}
\caption{
\label{aggregatorFirmFig} 
Overview of the aggregator board firmware.
It is composed of five parts: the trigger and readout management, the front-end interface, the Data Acquisition (DAQ), the Ethernet bridge, and the slow control.
}
\end{figure}

An overview of the aggregator board firmware is shown in Fig.~\ref{aggregatorFirmFig}.
It consists of five main blocks: the trigger and readout management, the front-end interface, the Data Acquisition (DAQ), the Ethernet bridge, and the slow control.

\subsubsubsection{Trigger and readout management}
The trigger and readout management is mainly designed to interface with the CRU through the GBT interface, consequently, it is used to synchronize the prototype with the CRU DAQ clock, to receive the trigger messages, and to permit data taking.
The GBT interface is derived from the GBT-FPGA core \cite{GBT_FPGA_paper} to be able to recover the clock from the reception link.
This recovered clock is then used to clock the GBT transmission side and the whole acquisition firmware thanks to the external and internal PLLs.
When the FoCal prototype is inserted in the readout system of the ALICE experiment, the GBT transmission side is used to transfer the readout data to the CRU in GBT packet mode (80 bits of data transmitted  at every 40\,MHz clock cycle).
On the reception side, the GBT is used to receive the trigger messages which are used (i) to request an event readout (physics trigger) and (ii) to provide the information needed to fill the header of the data packets produced by the electronics.
The header contents allow for unique identification of the event across the ALICE experiment.
The FoCal prototype can also be operated outside of the ALICE framework, i.e. without a GBT connection to a CRU.
In that case, the external PLL (see Fig.~\ref{aggregatorHardFig}) previously used in loopback mode and jitter cleaner, is used as a local clock source.
A `CTP emulator' is then utilized to produce trigger messages equivalent to those normally produced by the ALICE CTP and forwarded by the CRU.
This `CTP emulator', which is the same as the one implemented in the CRU, can be configured to produce periodic triggers or to use an external trigger input.
The `DAQ\_controller' then receives the trigger messages from the local or the remote source.
It assesses whether the trigger is sufficiently delayed with respect to the previous one (to cope with the HGCROC readout duration of 1.075\,\textmu s) and if there is enough buffer space available in the DAQ part to accept a new event.
A data drop request is produced if any of the two previous conditions are not fulfilled.
Finally, the `DAQ\_controller' sends to the front-end interface the appropriate data fields extracted from the trigger message and the drop request.
The `FCMD producer' is described in section~\ref{devKitFirmSect}.

\begin{figure}[hbtp]
\centering
\includegraphics[angle=0,width=0.99\textwidth]{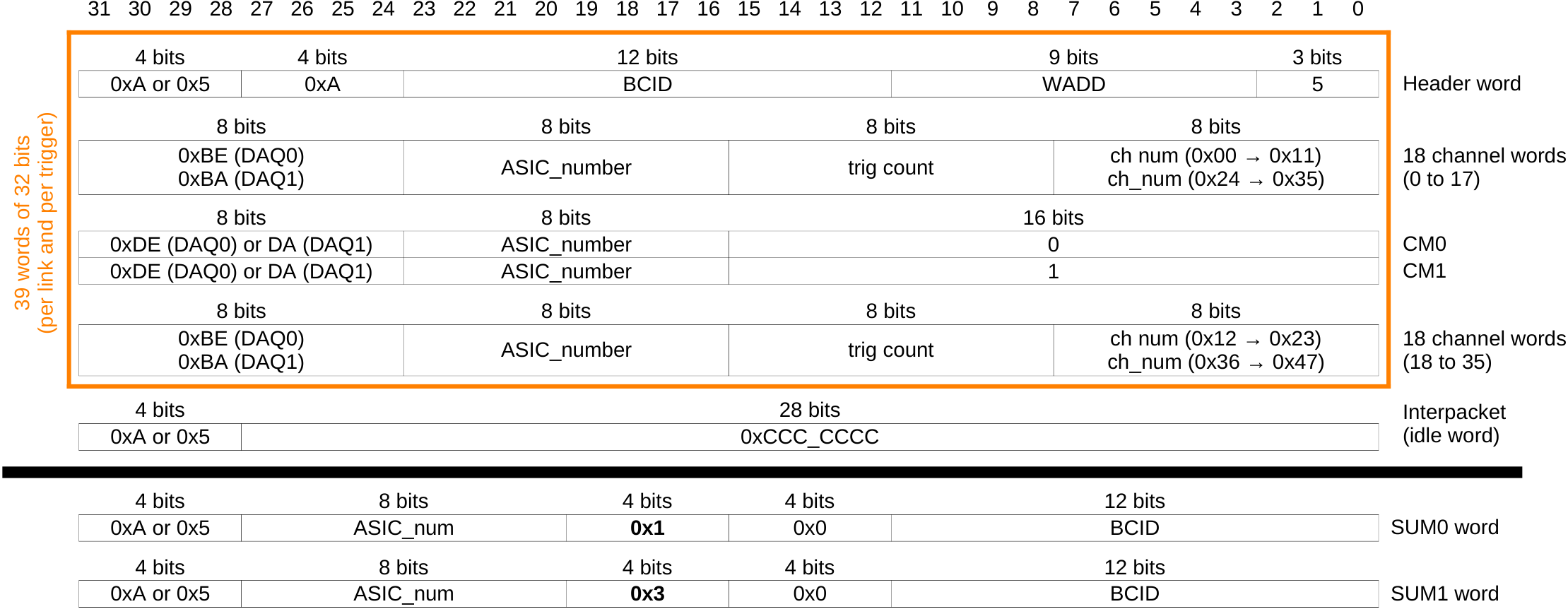}
\caption{
\label{hgcrocSimulFig} 
Description of the data produced by the HGCROC simulator component.
The top figure shows the simulated payload of 39 successive 32-bit words which is representative of those produced by the DAQ links of the HGCROC.
It features various water-markings to uniquely identify the data payload, its content, and the link simulated.
The bottom figure shows the 32-bit words produced at each machine clock cycle (40\,MHz) by each SUM link and their respective 4-bit marking.
}
\end{figure}

\subsubsubsection{The front-end interface} 
The front-end interface features 20 instances of the HGCROC interface component described in section~\ref{devKitFirmSect}.
In case any of the single-pad boards are not connected, or simply for firmware/software debug or test purposes, each `hgcroc\_interface' can be bypassed and replaced by an internal data generator that mimics the HGCROC behavior and produces unique and recognizable data for the DAQ and SUM links (see Fig.~\ref{hgcrocSimulFig}).
Each selected data source, featuring four 32-bit word buses transferring data at 40\,MHz, is then fed to `data reorganizer' components that perform several tasks at each trigger.
Each of them reorganizes the data of the two DAQ links and produces a payload of 39 64-bit words with the channel data organized in little endianness and with the common information shifted at the end of the payload (calibration cell data and common mode data).
This DAQ payload is stored in a 64-bit wide and 2048-word deep FIFO memory buffer.
In parallel, the data of the two SUM links are recorded for a duration determined by the configuration of the trigger and readout management block and stored in another FIFO of 64-bit wide and 2048 words deep.
Each FIFO can be read out by the DAQ block at a frequency of 120\,MHz.
Provided the SUM record duration is configured to less than or equal to 39 consecutive samples, the front-end interface has a buffering capability of about 52 events.
For monitoring purposes, each link is equipped with a 16-bit counter that counts each successful HGROC DAQ frame reception, and a latency meter that determines, for each link, the delay (in machine clock cycles) between the readout fast command and the deserialized DAQ payload reception.
Finally, the front-end interface contains a component in charge of storing all relevant trigger information required to build the final event payload, i.e. the Orbit number (32 bits), the bunch crossing ID (12 bits), the trigger type (32 bits) and the data drop information (1 bit).
This information is provided by the trigger and readout management block.

\subsubsubsection{Data Acquisition (DAQ) part}
The DAQ component's main tasks are to regroup the DAQ and SUM data and to build the data packets to be sent either via the Ethernet bridge or the GBT interface.
The DAQ and SUM data of each HGCROC are each regrouped in large FIFO buffers (32768 64-bit words) embedded in the `daq collector' components.
With a DAQ payload of 39 64-bit words and a typical SUM record size of 20 samples (two words of 32 bits) per HGCROC, a storage of 780 and 400 words, respectively, is required.
This results, at worst, in a buffering limit of 42 events.
These `daq collector' are filled in parallel (SUM and DAQ) at a clock rate of 120\,MHz which corresponds to an instantaneous input throughput of 960\,MBytes/s.
This throughput should be compared to the maximal data throughput produced by the 20 HGCROCs read out every 1.075\,\textmu s at best, i.e. 725\,MBytes/s.
The data packets are constructed to be compatible with the requirements of the readout through the CRU that limits the CRU data packet size to no more than 512 GBT words (header included).
Due to the GBT word odd size of 80 bits, this requirement yields a limitation of 512 64-bit words per data packet.
Consequently, with a total DAQ payload of 780 64-bit words and a typical total SUM payload of 400 words, the total payload amounts to 1180 64-bit words.
Distributing these data into three packets, and thus with three additional headers, the total event size is 1192 64-bit words.

The `packet builder' component is composed of a state machine, in charge of constructing the CRU data packets in accordance with the O2 protocol, and a FIFO buffer (32768 64-bit words) to store them before pushing them to the transmission interfaces (GBT or Ethernet bridge).
The state machine produces the periodic heartbeat frame start and stop packets (two packets of four GBT words) at every orbit of 89.1\,\textmu s duration and fills adequately the event data packet headers with the trigger messages information (Bunch Crossing Identification, orbit number, trigger bits, sequence number in the heartbeat frame).
When a data drop is requested, a small packet featuring the header and a single GBT word in the payload is produced and transferred to the transmission interfaces.
To fill the CRU data packets with the event data, the state machine, which operates at a clock rate of 120\,MHz, reads out the `DAQ collector' FIFO one by one and writes the resulting packets in the output FIFO.
Finally, depending on the use case (ALICE readout or standalone acquisition), the transmission interface can be achieved through GBT or Ethernet.

In GBT mode, a GBT word can be sent every 40\,MHz clock cycle.
To mark a data packet, two extra words must be sent too (start of packet and end of packet).
Consequently, sending the three data packets corresponding to one typical event packet of 1192 words takes 1198 clock cycles.
Then, accounting for the heartbeat start and stop packets (4\,+\,2 words each) emitted at every orbit (3564 machine clock cycles), the maximal achievable trigger rate is:
\begin{equation}
\rm \frac{3564-2 \times 6}{1198} \times \frac{40\,MHz}{3564} \sim 33\,kHz 
\end{equation}
This computed maximum trigger rate of this prototype readout does not yet match the expected rate in the experiment of about 1\,MHz, however in the final implementation the data payload will be significantly reduced, thanks to zero suppression, and the SUM information will not be recorded.
The `gbt packet xmit' blocks use an internal shallow FIFO buffer to accommodate the clock domain crossing and a state machine that inserts the required start and stop words around each packet.

In the User Datagram Protocol (UDP) mode, at the very best, a byte can be sent every 125\,MHz clock cycle.
This figure does not account for the encapsulation or the layer arbitration required by the successive layers: Ethernet, Address Resolution Protocol (ARP), Internet Protocol (IP), UDP, etc.
To improve the Ethernet layer usage, the `udp packet xmit' block is designed to use jumbo frames (Ethernet frames with more than 1500 bytes of payload) having a maximum size of 8192 bytes.
To do this, the `udp packet xmit' block tries to agglomerate successive CRU data packets produced by the `packet builder' layer in order to reach 8192 bytes.
To achieve this, it features a buffer space of 16,384 bytes and a management FSM.
This FSM ensures that a UDP frame is sent if one of the three following conditions is fulfilled: a UDP packet having a size of 8192 bytes is ready, there is not enough space available in the intermediate buffer to receive the next CRU data packet produced by the `packet builder`, or more than 100\,ms have elapsed since the last packet emission.

\subsubsubsection{Slow control part} 
The slow control part is composed of five low-level I2C masters controlling the external PLLs and the SFP modules, two monitor masters, and four high-level I2C masters for the HGCROCs.
All these components are managed via an Ethernet link through an IPBUS controller.
Each low-level I2C master implements an FSM to manage indirect I2C addressing, and hence to permit easy hardware access. 
The high-level masters, one per 5-pad-layer, are an evolution of the component described in section~\ref{devKitFirmSect}.
Each of these high-level masters addresses five HGCROCs and is equipped with a command buffer able to hold up to 512 commands.
Provided that a large number of commands is required to configure the HGCROCs at detector startup or between data taking, this buffer allows for the storage of the command list and having it executed one after the other without dead time.
Proceeding this way allowed for a substantial shortening of the initial HGCROC configuration time from 35 seconds down to three seconds.
As a consequence, the detector calibration runs last from several hours to a matter of a few tens of minutes, being compatible with data taking. 

The last component of the slow control is the `monitor master' which provides an interface between the IPBUS layer and the ten monitoring ADCs (ADS114S08).
This interface is used in two modes: the manual mode where each ADC internal 8-bit register can be accessed one by one with an IPBUS command, and the automated mode where the ten ADCs, organized in three different functional groups (voltages, currents, temperatures), are automatically controlled by an internal FSM.
Each functional group contains a dedicated FSM and a memory block of 64 16-bit words.
The FSM applies the proper sequence to switch the input channels, select the gain if applicable, start the conversions, read out the conversion result, and store it in memory.
The automated sequence is repeating itself periodically every 461\,\textmu s.
Consequently, the detector control software can read the detector health parameter at any time and get up-to-date values.

\subsubsubsection{Resource usage} 
The firmware described here uses 38,619 Look Up Tables (19\%), 58,787 Flip Flops (14\%), and 296 Block RAMs (54\%).
Consequently, this leaves a significant margin to allow for the implementation of additional functionalities such as zero suppression algorithm and local trigger information construction.

\subsection{LED driver board}
\begin{figure}[hbtp]
\centering
\includegraphics[angle=0,width=0.49\textwidth]{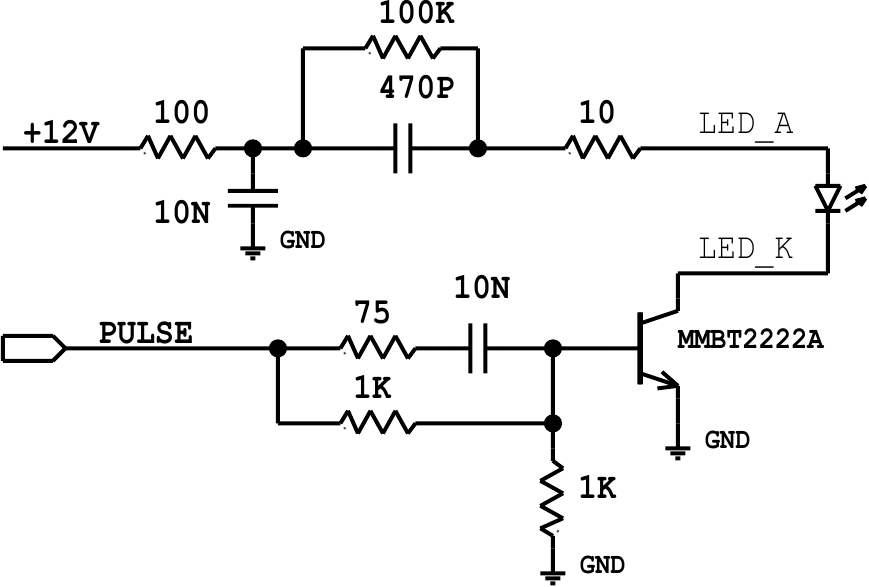}
\includegraphics[angle=0,width=0.49\textwidth]{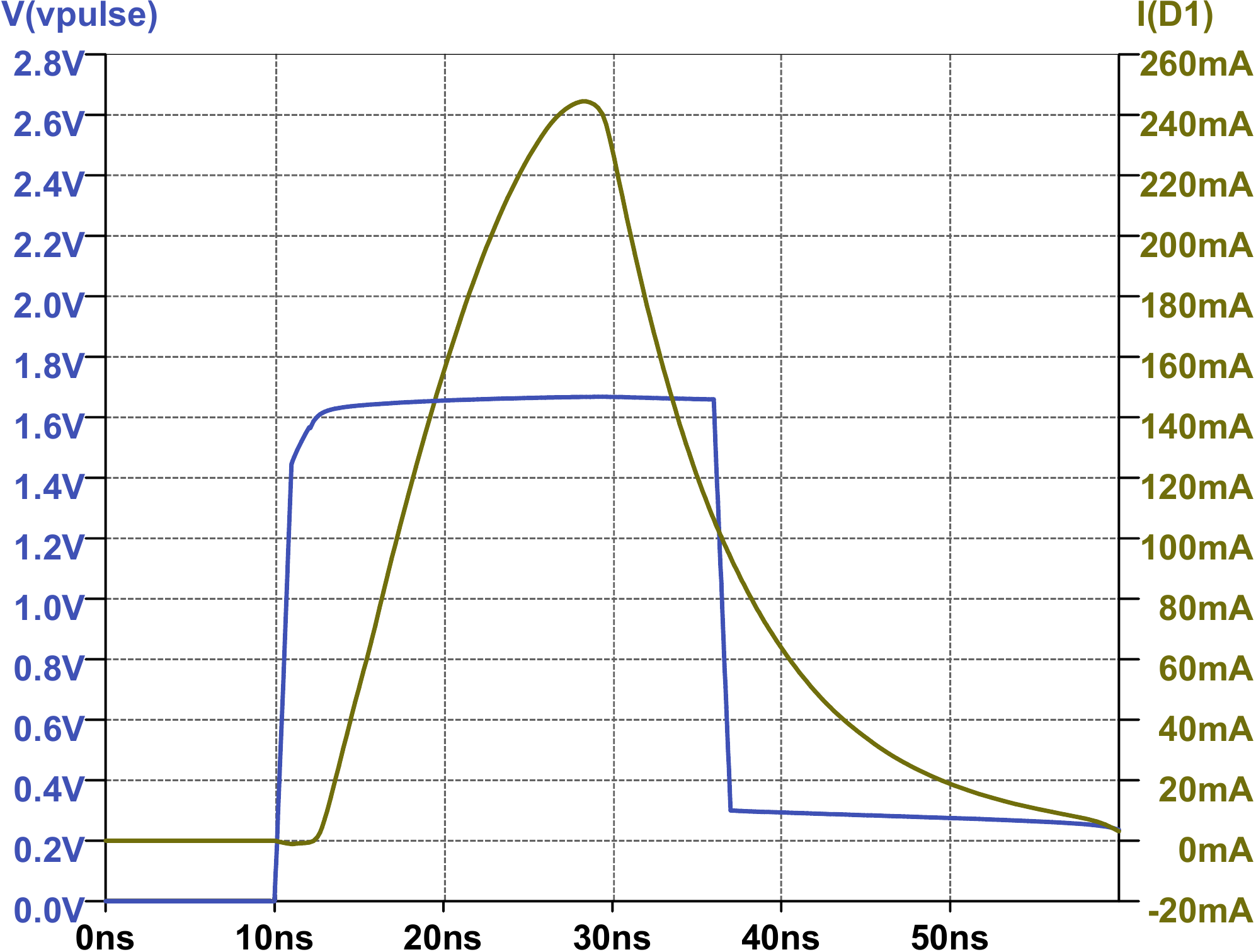}
\caption{
\label{ledSimuFig}
Schematic of the LED driver on the left, and SPICE simulation of the circuit on the right.
The simulation shows the current peak value and shape for a typical control signal of one machine clock period of 25\,ns.
}
\end{figure}

A light injector was developed to test and validate the successful assembly and wire bonding of silicon pad sensors onto single pad boards.
The  light injector, controlled by the trigger output of the KCU105 development kit or the aggregator, is composed of an amplifier circuit and an Infra-Red (IR) Light Emitting Diode (LED) from Vishay\texttrademark{} (VSMY14940), see Fig.~\ref{ledSimuFig}. 
The amplifier circuit was designed to generate a sharp current pulse to produce as much light as possible while keeping a duration similar to the HGCROC sampling period.
Due to the lack of an appropriate model, the simulation was conducted with a red LED (QTLP690C model) having a higher forward voltage.
It was verified that the maximum current and the duration are mostly dependent on the transistor gain and on the control pulse voltage but it is slightly dependent on the diode type (about 25\,mA of peak current variation for a forward voltage variation between 1.9\,V to 3.38\,V).

Given the opacity of the gold-plated free side of the sensor (opposite to the glued one), this single-channel light injector was used to inject light from the component side of the single-pad board through the PCB holes.
In anticipation of these tests, the wire bonds and the cavities were covered with an IR-translucent glop top.

The light injector is also used for timing adjustment between the trigger signal and the HGCROC readout command, in particular for cosmic ray tests of very low trigger rates and during beam tests with limited access to the experimental area.
To achieve this, a dual-output pulse generator is used to fire the light injector and, after a delay corresponding to the light injector latency, feeds the trigger logic used by the readout board.
Then, either the HGCROC internal circular buffer delay (negative) or the delay added by the `trigger generator and conditioner' (positive) are adjusted to scan the timing adjustments and ensure that the optimal delay was calculated.

\begin{figure}[hbtp]
\centering
\includegraphics[angle=0,width=0.49\textwidth]{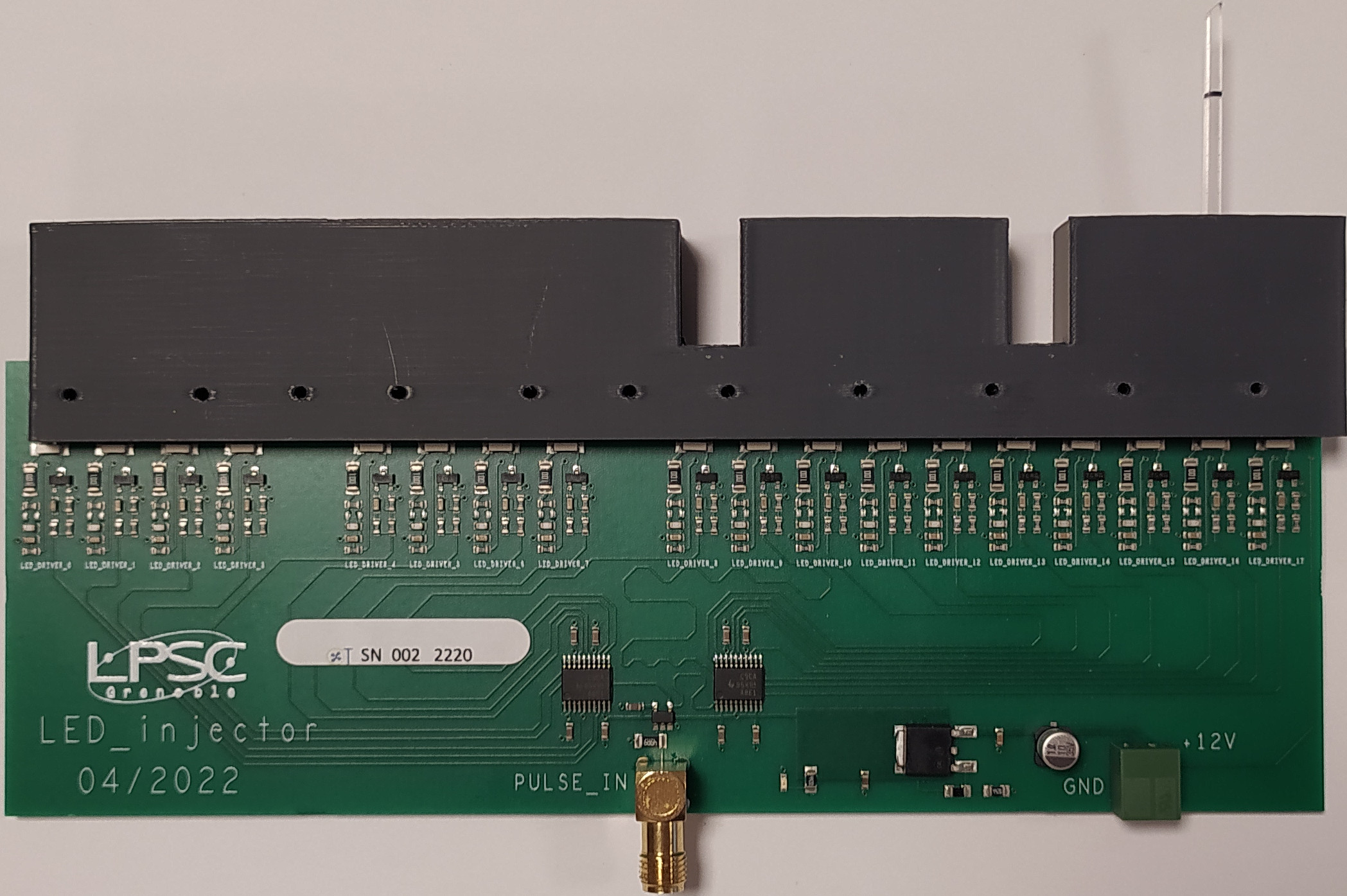}
\includegraphics[angle=0,width=0.49\textwidth]{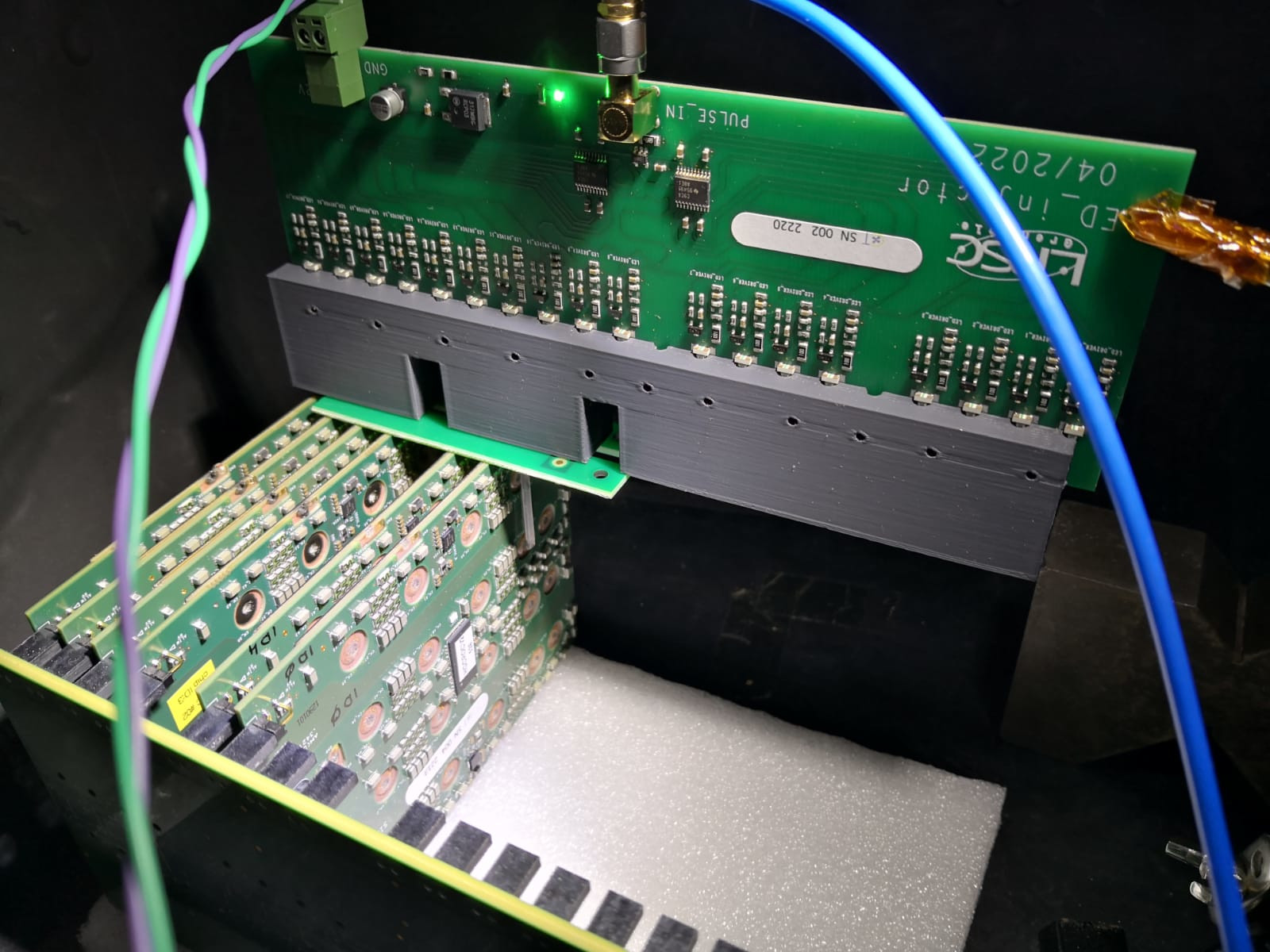}
\caption{
\label{multiInjFig} 
Pictures of the multi-channel light injector top view shown on the left and assembled with five single pad boards on the right.
An optical fiber support was designed and 3D printed, it is used to hold the optical guides in place (only one inserted here) and allows an easy assembly with the prototype.
}
\end{figure}

To be able to inject light, after assembly, in at least one cavity of each single-pad board of the prototype tower, a multi-channel light injector was designed and produced (shown in Fig.~\ref{multiInjFig}) implementing 18 elementary light injectors.
An optical fiber support was designed and 3D printed, to hold the optical guides in place, and it allows easy mounting on the prototype.
The 18 light injectors are controlled from a single input signal distributed through two low-skew buffers (CDCLVC1110), and thus applied simultaneously to all channels.

\section{Conclusion}

A prototype readout electronics was designed to equip the FoCal pad detector planned as an ALICE upgrade for the LHC Run 4. 

The very front-end `on-detector' electronics embeds the CMS HGCal readout HGCROC chip which measures and digitizes the charge deposited in the silicon pad sensors.
The front-end board PCBs are drilled with through-holes for the wire bonding of the sensors.
The layout of this 10-layer PCB and the optimization of its cavities are described in this publication.

The FPGA-based back-end electronics is in charge of controlling, reading out the HGCROCs, and shipping the data to CRUs through GBT links.
Various tools have been developed to validate the readout electronics and the operation of the detector before (charge injector board) and after (LED driver) the wire bonding of the pad sensors. 

A fully instrumented FoCal electromagnetic calorimeter tower equipped with the prototype readout presented in this publication was built and operated in CERN test beam facilities \cite{Rauch_TB}. 
The prototype readout electronics successfully proved to meet its design goals.

\section*{Acknowledgments}
We thank Franck AGNESE, Olivier CLAUSSE, and Christophe WABNITZ (C4Pi platform from IN2P3-IPHC laboratory) for their insight on the required PCB optimizations to permit wire bonding and for all the operations required to ensure good sensor assemblies (wire bonding, sensor grounding, glop top encapsulation, and 3D measurements).

We are also grateful to Christophe HOARAU from LPSC for the thermal simulations conducted on the 5-pad-layer board.

And finally, we would like to thank the ALICE FoCal Collaboration in which the work presented in this publication was undertaken. 

\bibliography{focal_bibtex}
\bibliographystyle{JHEP}

\end{document}